\documentclass[a4paper,11pt]{article}
\pdfoutput=1
\usepackage{jheppub}
\usepackage[T1]{fontenc}

\title{\boldmath Non-Conformality, Subregion Complexity and Meson Binding}
\author[a]{Mahsa Lezgi,}
\author[a]{Mohammad Ali-Akbari}
\author[b]{and Mohammad Asadi}
\affiliation[a]{Department of Physics, Shahid Beheshti University, Tehran, Iran}
\affiliation[b]{IPM, School of Particles and Accelerators, P.O. Box 19395-5531, Tehran, Iran}
\emailAdd{s\_lezgi@sbu.ac.ir}
\emailAdd{m\_aliakbari@sbu.ac.ir}
\emailAdd{m\_asadi@ipm.ir}

\abstract{We study holographically the zero and finite temperature behavior of the potential energy and holographic subregion complexity corresponding to a probe meson in a non-conformal model. We observe that in zero and low temperature non-conformality has a decreasing effect on the dimensionless meson potential energy. However, non-conformal corrections increase absolute value of the dimensionless holographic subregion complexity in both zero and finite temperature which means the non-conformal state needs less information to be specified. In other words, considering the effect of non-conformality, the less bounded meson state needs less information to be specified. In low temperature limits, thermal corrections decrease meson potential energy and do not have a specific effect on holographic subregion complexity. We find that in the vicinity of the phase transition, the zero temperature meson state is more favorable than the finite temperature state, from the holographic subregion complexity point of view.}

\begin{document} 
\maketitle
\flushbottom
\section{Introduction}
The gauge/gravity duality is a conjectured relationship between quantum field theory and gravity. This duality is a strong-weak duality which maps a strongly-coupled quantum gauge field theory to a weakly-coupled classical gravity in a higher dimension \cite{maldacena}. The duality provides an important framework to study key properties of the boundary field theory dual to some gravitational theory on the bulk side. This idea is used to describe phenomena in strongly-coupled quantum field theories ranging from condensed matter physics to low- energy quantum chromodynamics (QCD), the theory of the strong interactions \citep{mateos}. The most significant example of gauge/gravity duality is the AdS/CFT correspondence which proposes a duality between IIB string theory on AdS spacetimes in $d+1$ dimensions and $d$-dimensional superconformal field theories. This framework has been applied to study quantities such as Wilson loops, entanglement entropy and recently extended to the quantum computational complexity in field theory. The generalization of AdS/CFT correspondence to field theories which are not conformal seems to be important. It is then interesting to develop our understanding of this duality for more general cases.

There are many different families of non-conformal field theories and one can study the effect of the non-coformality on their physical observables \cite{thermo}. One of the most important gauge-invariant and non-local observables in gauge theories is Wilson loops. In quantum field theory, potential energy can be obtained by this quantity as a gauge invariant non- local operator. The expectation value of Wilson loops has valuable applications to the confinement/deconfinement and QCD-like theories \cite{quark,quark2,erdmenger1,erdmenger2}. In particular the expectation value of this operator on a rectangular loop gives us the potential energy between a static quark and anti-quark.

Entanglement entropy is another intriguing non-local quantity which measures the quantum entanglement between two sub-systems of a given system. This quantity is extremely useful in many quantum systems, ranging from condensed matter physics to black hole physics. It can be used to classify the various quantum phase transitions and critical points \cite{lezgi1,kleb,kleb2}. Since the quantum field theories have infinite degrees of freedom, the entanglement entropy is divergent. Thus, it is a scheme-dependent quantity and needs to be regulated. It has been shown that the leading divergence term is proportional to the area of the entangling surface (for $d>2$) \citep{kleb3,kleb4}. The entanglement entropy has a holographic dual given by the area of minimal surface extended into the bulk whose boundary coincides with the boundary of the subregion \cite{takayanagi}.  

Concepts from quantum information theory are having a rapidly growing influence in investigations of quantum field theory and quantum gravity. One of the most important information quantities is quantum computational complexity. This is a state-dependent quantity which measures how difficult is to prepare a given state, from the reference state. More precisely, quantum computational complexity involves minimizing the number of unitary transformations required to transform a reference state to a target state. For more details, see \cite{john,scott}. There have been two prescriptions to calculate the quantum complexity in terms of the gauge/gravity duality, which are known as the CV (Complexity=Volume) conjecture \cite{susskind1} and the CA (Complexity=Action) conjecture \cite{susskind2,susskind3}.  Note that these proposals correspond to the complexity of a pure state in the whole boundary space of the dual quantum field theory. If one would like to compute the complexity of a mixed state, corresponding to a subsystem, then one should generalize the notion of complexity to the subregion complexity. There are two known proposals called holographic subregion complexity (HSC) for the CV conjecture \cite{Alishahiha} and the CA conjecture \cite{comments}. The reader can find other studies on the holographic subregion complexity in \cite{quench,Zhang:2017,sub:2019,volume,Zhang:2019,renormalization,asadi}.

The remainder of the paper is organized as follows: Section \ref{aa} considers a brief review of modified AdS$_5$ (MAdS) background and its black hole version, which is called modified black hole (MBH) and we investigate the thermodynamics properties of them including entropy density and pressure, which are given in terms of non-conformal parameter of theory. In section \ref{bb}, we study some non-local observables such as potential energy, entanglement entropy and HSC corresponding to a bound state meson and express how to calculate them through holographic prescriptions. In section \ref{cc} we study the meson potential energy at both zero and finite temperatures. In order to obtain the analytical results we develop a systematic expansion, see \cite{fischler,sarkar,ebrahim}, up to some specific order of the expansion parameter and consider a specific limit of the underlying field theory called high energy limit. However, since we can not obtain analytical results, we will not study the low energy limit. At finite temperature, we focus on two favorable low and high temperature limits. In section \ref{dd}, we go through the same steps as section \ref{cc} and obtain the analytical results corresponding to the HSC. In section \ref{dd6}, we conclude with a brief discussion of our results and we consider some directions for future research. In appendix \ref{A}, we present the full details of our calculations corresponding to meson potential energy. In appendix \ref{B} we do the same for HSC results.   
\section{Review on the backgrounds}\label{aa}
We are interested in studying the non-local quantities such as Wilson loop and HSC of a probe meson (a stable quark and anti-quark bound state) using the framework of holography. Therefore, we start with a five-dimensional background and its black hole version, which is called modified AdS$_5$ (MAdS) and modified black hole (MBH), respectively. These backgrounds are dual to QCD-like gauge theories at zero and non-zero temperature.
\subsection{The backgrounds}
The gravitational backgrounds we study here are the following, MAdS background \cite{Andreev1}
\begin{eqnarray}\label{1}
ds^2&=\frac{r^2}{R^2}g(r)\left(-dt^2+d\vec{x}^2+\frac{R^{4}}{r^4}dr^2\right),\,\,\,\,\,\,\,\,\,g(r)=e^{\frac{cR^{4}}{2r^2}},
\end{eqnarray}
and the black hole version of the above model, MBH background, is introduced by  \cite{Andreev2} 
\begin{align}
ds^2&=\frac{r^2}{R^2}g(r)\left(-f(r)dt^2+d\vec{x}^2+\frac{R^{4}}{r^4f(r)}dr^2\right),\,\,\,\,\,\,\,\,\,\,\,\,\,\,\,f(r)=1-\frac{r_h^4}{r^4},
\label{2}
\end{align}
where $r$ is the radial coordinate, $r_h$ is the position of the horizon, $\vec{x}\equiv (x,y,z)$, $R$ is the asymptotic AdS$_5$ radius and the boundary is located at $r=\infty$ where QCD-like model lives.
The modifier parameter $c$ has (energy)$^2$ dimension whose value is fixed from the $\rho$ meson trajectory  estimated to be of order $0.9$ GeV$^2$ \citep{Andreev2}. There is an upper bound on the maximum value of the radial coordinate called $r_c =R^2\sqrt{\frac{c}{2}}$. Note that as $c$ goes to zero then $r$ is not bounded, as expected for the AdS background \cite{Andreev1}. In the QCD-like theory, $r_c$ corresponds to a mass gap which is the lowest excitation and its energy order is the same as the QCD scale. Therefore, we assign a energy scales $\Lambda_c$ to $r_c$ in such a way that  $\Lambda_{c}\equiv\sqrt{c}$. The asymptotically AdS$_5$ solution and and its black hole version will be respectively recovered if one sets the modifier parameter zero, say $c=0$, in backgrounds \eqref{1} and \eqref{2}.
\subsection{Thermodynamics}
Consider a general class of black brane metrics of the form
\begin{equation}
ds^2=g(r)[-f(r)dt^2+d\vec{x}^2]+\frac{1}{h(r)}dr^2,
\label{metric1}
\end{equation}
where $f(r)$ and $h(r)$ have a first order zero at the horizon $r=r_h$, while $g(r)$ is non-vanishing there. The Hawking temperature $T$ and the entropy density $s$ of the black brane  is given by
\begin{equation}
T=\frac{\sqrt{g(r)f'(r)h'(r)}}{4\pi}\bigg{|}_{r=r_h},\qquad  s=\frac{g(r)^{\frac{3}{2}}}{4G_N}\bigg{|}_{r=r_h},
\label{TS1}
\end{equation}
where $G_{N}$ is the Newton constant. According to \eqref{2} and using the fact that $f(r_h)=0$, the temperature and the entropy density of the boundary field theory become
\begin{equation}
T=\frac{r_h}{R^2\pi},\qquad \qquad \qquad \qquad  \qquad  s=\frac{r_h^3\,\,e^{\frac{3\Lambda_c ^2 R^4}{4r_h^2}}}{R^3 4G_N}.
\label{TS2}
\end{equation}
It is clearly seen from the above equations that the theory will be conformal if one takes the limit $c\rightarrow 0$ (or $\Lambda_c \rightarrow {0}$). Hence, one expects that the entropy density must coincide with that of a relativistic conformal theory which scales as $T^{3}$. We are now in a position to determine the pressure $p$ and the energy density $\epsilon$ of the thermal system through standard thermodynamic relations  
\begin{equation}
p=\int dT\,\, s(T),\qquad \qquad \qquad \qquad  \qquad  \varepsilon +p=Ts.
\label{PE1}
\end{equation}
Using \eqref{2}, \eqref{TS2} and \eqref{PE1} the pressure and the energy density are expressed
\begin{align}
&p=\frac{\pi^ 3 R^3}{4G_N}\bigg[e^{\frac{3\Lambda_c ^2}{4\pi^ 2 T^2}}\bigg(\frac{3\Lambda_c ^2T^2 }{16\pi ^2}+\frac{T^4 }{4}\bigg)-\frac{9 \Lambda_c ^4 \,Ei(\frac{3\Lambda_c ^2}{4\pi^2 T^2})}{64\pi ^4}\bigg], 
\nonumber\\&\varepsilon=\frac{\pi^ 3 R^3}{4G_N}\bigg[e^{\frac{3\Lambda_c ^2}{4\pi^ 2 T^2}}\bigg(\frac{3T^4 }{4}-\frac{3\Lambda_c ^2T^2 }{16\pi ^2}\bigg)+\bigg(\frac{9 \Lambda_c ^4 \,Ei(\frac{3\Lambda_c ^2}{4\pi^2 T^2})}{64\pi ^4}\bigg)\bigg],
\label{PE2}
\end{align}
where $Ei(x)$ is the exponential integral  defined for a real non-zero values of $x$
\begin{eqnarray}
Ei(x)=-\int_{-x}^{\infty} dt \frac{e^{-t}}{t}.
\end{eqnarray}
\section{Wilson loop, entanglement entropy and subregion complexity}\label{bb}
We explore the behavior of some non-local observables including potential energy and sub- region complexity of a probe meson. To obtain analytic expressions for these observables we do a systematic expansion at both zero and finite temperature using the holography idea.
\subsection{Wilson loop}
In quantum filed theory, the potential energy can be obtained by the Wilson loop operator. The expectation value of this operator on a rectangular loop, $\cal{R}$ with two sides, time $\tau$ and distance $l$, where $\tau\gg l$, gives us the potential energy between a static quark and anti-quark with the distance $l$. The holographic prescription of the expectation value of Wilson loop is the on-shell action of a classical string $S(\cal{R})$ whose endpoints correspond to quark and anti-quark on the boundary and is suspended from the boundary to the extra dimension in the bulk \cite{wilson}
\begin{eqnarray}
\langle W({\cal{R}})\rangle =e^{i S({\cal{R}})}.
\label{wilson}
\end{eqnarray}
It is then straightforward to find the potential energy corresponding to the binding energy
of the meson. Further information about holographic Wilson loop can be found in \cite{erdmenger2}.
\subsection{Entanglement entropy}
If we decompose the total Hilbert space of a system, $\mathcal{H}_{tot}$, into two subsystems $\mathcal{H}_{A}$ and $\mathcal{H}_{B}$ such that $\mathcal{H}_{tot}=\mathcal{H}_{A}\otimes \mathcal{H}_{B}$, then we will trace out the sub-system $B$ and define the reduced density matrix $\rho_A$ for the subsystem $A$ as $\rho_A=Tr_B\rho$ where $\rho$ is the total density matrix. The entanglement entropy $S_{A}$ measures how much information is hidden inside the sub-system $A$ and defined as the Von Neumann entropy of the reduced density matrix $\rho_{A}$
\begin{eqnarray}
S_{A}=-tr \rho_{A}\log\rho_{A}.
\end{eqnarray}
The AdS/CFT correspondence provides an elegant way to compute the entanglement entropy in terms of a geometrical quantity on the bulk. This is the so-called holographic entanglement entropy formula, first proposed by Ryu and Takayanagi \cite{takayanagi,Ryu}
\begin{eqnarray}\label{RT}
S_{A}=\frac{Area(\gamma_{A})}{4G_{N}^{d+2}},
\end{eqnarray}
where $S_{A}$ is the holographic entanglement entropy for the sub-system $A$, $\gamma_{A}$ is a codimension-two minimal area surface (RT surface) whose boundary $\partial \gamma_{A}$ coincides with $\partial A$, and $G_{N}^{d+2}$  is the $d+2$-dimensional Newton constant.
\subsection{Subregion complexity: CV duality}
If we would like to compute the complexity of the mixed state, then we will need to extend the holographic complexity to the subregions. The holographic volume prescription for cal- culating subregion complexity states that the HSC for a subregion $A$ on the boundary equals the volume of codimension-one RT surface enclosed by $\gamma_{A}$ which is given by the following form \cite{Alishahiha}
 \begin{eqnarray}
{\cal{C}}_A=\frac{V_{\gamma_{A}}}{8\pi RG_N} ,
\label{CV}
\end{eqnarray}
where $R$ is AdS radius and ${\cal{C}}_A$ is known as the HSC for the subregion $A$.
\section{Potential energy}\label{cc}
Heavy-quark potential is one of the fundamental observables which is relevant to confinement. Based on \cite{lattice}  the following heavy-quark potential, Cornell potential, is proposed
\begin{eqnarray}
V(r)=-\frac{\kappa}{r}+\frac{r}{a^2}+\cal{V},
\label{cornell}
\end{eqnarray}
where  $\kappa$ is treated as a phenomenological parameter and $a$ is inferred from lattice gauge theory. These parameters are adjusted to be 
\begin{eqnarray}
\kappa \approx 0.48, \qquad a\approx 2.34 GeV^{-1}, \qquad  {\cal{V}} \approx -0.25 GeV.
\end{eqnarray}
The first term of the potential is in complete accordance with the known Coulomb potential at short distance $r\rightarrow 0$ and the second one with confinement at large distance $r\rightarrow\infty$.

At zero temperature, in the background \eqref{1}, the potential energy between the quark and anti-quark pair is obtained and shown that this background describes the low energy of QCD-like theory, confined phase \cite{Andreev1}. This meson potential energy is given by 
\begin{align}\label{pot} %
V(r)=\left\{%
\begin{array}{ll} %
p\left(-\frac{\kappa_0}{r}+\sigma_0 r+ O(r^3)\right), \ \ \ \ \ r\rightarrow 0 \\
p(\sigma r), \ \ \ \ \ \ \ \ \ \ \ \ \ \ \ \ \ \ \ \ \ \ \ \ \ \ \ \ r\rightarrow\infty\\
\end{array}%
\right.
\end{align} %
where $p\approx0.94$ and $\kappa_0\approx0.23$ are dimensionless parameters, and we have $\sigma_0\approx0.16$ GeV$^2$ and $\sigma\approx0.19$ GeV$^2$ for $c=0.9$ GeV$^2$. These constants are fixed according to the Cornell potential \eqref{cornell}. In the following subsections we would like to obtain analytically the potential energy of the probe meson at both zero and finite temperatures. In the rest of the paper we set the AdS radius $R=1$. 
\subsection{Potential energy: Zero temperature expansion}\label{cc1} 
In this section we consider a state which corresponds to the MAdS background \eqref{1} and probe this state with a meson in the QCD-like theory. On the gravity side, this meson is dual to a classical string whose dynamic is given by Nambu-Goto action
 \begin{eqnarray}
S_{NG}=\frac{-1}{2\pi \alpha'}\int d\tau d\sigma \sqrt{-\det(g_{ab})} ,
\label{action}
\end{eqnarray}
where $\alpha'=l_{s}^{2}$ in which $l_s$ is the string length and $g_{ab}$ is the induced metric on the world-sheet. The world-sheet can be parameterized by $\tau=t$ and $\sigma=x$ and by demanding that $\tau\rightarrow\infty$, its shape is given by $r(x)$. We set the quark and anti-quark at $x=-l/2$ and $x=l/2$. At zero temperature using MAdS background \eqref{1} we can easily see that  
\begin{equation}
S_{NG}=\frac{\tau}{\pi\alpha'}\int_{0}^{\frac{l}{2}}e^{\frac{c}{2r^2}}r\sqrt{\frac{r'^2}{r^2}+r^2}dx,
\label{v1}
\end{equation}
where $r'=dr/dx$. The above action is not explicitly dependent on $x$ so the corresponding Hamiltonian is a constant of motion. Hence, using the boundary condition  $r'(x)\vert_{x=0}=0$, we get
\begin{equation}
r'(x)=\frac{r^2}{r_*^2}e^{\frac{-c}{2r_*^2}}\sqrt{r^4e^{\frac{c}{r^2}}-r_*^4e^{\frac{c}{r_*^2}}},
\label{v2}
\end{equation}
where $r_*=r(x)\vert_{x=0}$ is the returning point of the string. By integration of equation \eqref{v2} the characteristic length $l$ corresponding to the separation of quark and anti-quark read as follows 
\begin{align}
l(r_*)&=2r_*^2e^{\frac{c}{2r_*^2}}\int_{r_*}^{\infty}\frac{dr}{r^2\sqrt{•r^4e^{\frac{c}{r^2}}-r_*^4e^{\frac{c}{r_*^2}}}}
\cr &=\frac{2}{r_*}\int_{0}^{1}u^2e^{\left(\frac{r_c}{r_*}\right)^{2}(1-u^2)}  (1-e^{2(\frac{r_c}{r_*})^2(1-u^2)}u^{4})^{-\frac{1}{2}}du,
\label{v3}
\end{align}
where $u=r_*/r$. By substituting \eqref{v2} in \eqref{v1} and using \eqref{wilson} we have the following expression for the meson potential energy 
\begin{equation}
V_{q\bar{q}}(r_*)=\frac{r_*}{\pi \alpha'}\int_{\delta}^{1}u^{-2}e^{\left(\frac{r_c}{r_*}\right)^{2}u^2}\left(1-e^{2\left(\frac{r_c}{r_*}\right)^{2}(1-u^2)}u^{4}\right)^{-\frac{1}{2}}du,
\label{v6}
\end{equation}
where $\delta$ is an ultra violet cut off due to the  UV divergence structure of the potential energy.

Unfortunately \eqref{v6} can not be analytically solved. Therefore, we use binomial expansion, $\left(1-x\right)^{-\frac{1}{2}}=\displaystyle\sum_{n=0}^{\infty}\frac{\Gamma(n+\frac{1}{2})}{\sqrt{\pi}\Gamma(n+1)}x^{n}, \,\,-1\leq{x}<1$. 
By defining $x=e^{2\left(\frac{r_c}{r_*}\right)^{2}(1-u^2)}u^{4}$ and using the fact that $r_c<r_*$ one can easily see that $x<1$ and hence the sum is well-defined.
Following the above discussion, \eqref{v3} and \eqref{v6} can be indicated by the infinite series
\begin{align}
&l(r_{*})=\frac{2}{r_*}\sum_{n=0}^{\infty}\frac{\Gamma(n+\frac{1}{2})}{\sqrt{\pi}\Gamma(n+1)}\int_{0}^{1}u^{4n+2}~e^{(2n+1)(1-u^{2})\left(\frac{r_c}{r_*}\right)^{2}}du, \label{vv4} \\&
V_{q\bar{q}}(r_*)=\frac{r_*}{\pi \alpha'}\sum_{n=0}^{\infty}\frac{\Gamma(n+\frac{1}{2})}{\sqrt{\pi}\Gamma(n+1)} \int_{\delta}^{1}u^{4n-2}~e^{(2n+(1-2n)u^{2})\left(\frac{r_c}{r_*}\right)^{2}} du.
\label{v4}
\end{align}
In order to find the meson potential energy $V_{q\bar{q}}$ as a function of the characteristic length $l$ the following simple procedure is done. We should solve equation \eqref{vv4} for $r_*$ and then substitute it in equation \eqref{v4} to obtain $V_{q\bar{q}}$ in terms of $l$. In practice, we can not analytically solve equation \eqref{vv4} to find $r_*$ as a function of $l$. Therefore, we need to focus on the specific limit, which we call the high energy limit. We will introduce this limit in the next part. Note that, in the low energy limit, i.e, $r_{*}\rightarrow r_c{}$ (or $l\Lambda_{c}\gg 1$), we do not reach the analytical results and then we have to neglect studying this limit.
\subsubsection{High energy limit}\label{hev}
As we mentioned at the end of the section \ref{cc1}, to find $V_{q\bar{q}}$ in terms of $l$ we focus on the high energy limit. On the gravity side, by this limit we mean that the upper bound on the maximum value of the radial coordinate $r_c$ should be very smaller than the turning point of the classical string $r_{*}$, i.e. $r_{c} \ll r_{*}$. On the field theory side, the energy scale which we assign to $r_{c}$, called $\Lambda_{c}$, should be very smaller than the energy scale corresponding to the probe meson, i.e. $l\Lambda_{c}\ll 1$. It is noticed that, in this limit, the corrections to boundary observables are small and hence we can perturbatively do the calculations.

In high energy limit, from equation \eqref{vv4} and keeping up to the 4th order in $r_{c}/r_{*}$ we obtain
\begin{equation}
l(r_*)=\frac{2}{r_*}\bigg[a_1+a_2\left(\frac{r_c}{r_*}\right)^{2}+a_3\left(\frac{r_c}{r_*}\right)^{4} \bigg],\,\,\,\,\,\,\,\,\,\,\,
a_1,a_2,a_3>0,
\label{v8}
\end{equation}
where numerical coefficients $a_1$, $a_2$ and $a_3$ are reported in table \ref{list1} in the appendix \ref{A}. Solving equation \eqref{v8} perturabtively for $r_*$ and then considering the finite terms, the terms which do not include $\delta$ in the meson potential energy (the details of the calculations are written in the appendix \ref{A}), we finally reach the following expression
\begin{equation}
\tilde{V}_{q\bar{q}}(l\Lambda_{c})\equiv\frac{\alpha' V_{q\bar{q}}(l,l\Lambda_{c})}{\Lambda_{c}}=\frac{1}{l\Lambda_{c}}\bigg(\bar{b}_{1}+\frac{\bar{b}_{2}}{2}(l\Lambda_{c})^{2}+\frac{\bar{b}_{3}}{4}(l\Lambda_{c})^{4} \bigg),\,\,\,\,\,\bar{b}_{1}<0,\,\,\,\, \bar{b}_{2}, \bar{b}_{3}>0,
\label{vv12}
\end{equation}
where $\tilde{V}_{q\bar{q}}$ is the dimensionless meson potential energy at zero temperature and numerical coefficients $\bar{b}_{1}$, $\bar{b}_{2}$ and $\bar{b}_{3}$ are shown in table \ref{list1} in the appendix \ref{A}. Since the underlying field theory is conformal, it is expected that the dimensionless parameter $l\Lambda_{c}$ will appear. To compare correctly the underlying field theory with that of including the energy scale $\Lambda_{c}$, we make the meson potential energy dimensionless. Besides, by doing so we can get meaningful and intuitive results by taking the conformal limit $l\Lambda_{c}\rightarrow 0$. The first negative term in equation \eqref{vv12} is just related to the well-known Coulomb potential and the second and third positive terms come from the non-conformality appearance. The interesting point is that in high energy limit, $l\Lambda_{c}\ll 1$, by increasing $l\Lambda_{c}$ one can deduce that $\tilde{V}_{q\bar{q}}$ becomes less bounded, i.e. the more non-conformality effects, the less bounded dimensionless meson potential energy.
\subsection{Potential energy: Finite temperature expansion}\label{fte}
In this section we would like to study the thermal physics of meson potential energy $V_{q\bar{q}}$ in the MBH background \eqref{2}. We use the systematic expansion as we did in the previous section in both low temperature i.e. $lT \ll 1$ (or $r_{h}\ll r_{*}$) and high temperature i.e. $lT \gg 1$ (or $r_{h}\sim r_{*}$) limits. Using \eqref{2}, \eqref{action} and following the same previous calculations we reach the following expressions for $l$ and $V_{q\bar{q}}$
\begin{align}
l(r_*)&=\frac{2}{r_*}\int_{0}^{1}\frac{\sqrt{f(r_*)}}{f(u)}u^2e^{\left(\frac{r_c}{r_*}\right)^{2}(1-u^2)}\left(1-\frac{f(r_*)}{f(u)}e^{2\left(\frac{r_c}{r_*}\right)^{2}(1-u^{2})}u^{4}\right)^{-\frac{1}{2}}du,\label{vt0} \\
V_{q\bar{q}}(l)&=\frac{r_*}{\pi\alpha'}\int_{\delta}^{1}u^{-2}e^{\left(\frac{r_c}{r_*}\right)^{2}u^{2}}\left(1-\frac{f(r_*)}{f(u)}e^{2\left(\frac{r_c}{r_*}\right)^{2}(1-u^{2})}u^{4}\right)^{-\frac{1}{2}}  du,\label{vt1}
\end{align} 
where $\delta$ is an ultraviolet cut off and $f(u)=1-(r_h/r_*)^{4}u^{4}$. The above integrals can not be solved analytically, so we develop a systematic expansion by using binomial expansion. If we define $x\equiv(f(r_*)/f(u))e^{2\left(\frac{r_c}{r_*}\right)^{2}(1-u^{2})}u^{4}$,  we will then get a convergent series, following the fact that $r_c<r_{*}$ and $r_{h}<r_{*}$. Now, \eqref{vt0} and \eqref{vt1} are given by 
\begin{align}
l(r_{*})&=\frac{2}{r_*}\sum_{n=0}^{\infty}\frac{\Gamma(n+\frac{1}{2})}{\sqrt{\pi}\Gamma(n+1)}f(r_*)^{n+\frac{1}{2}}\int_{0}^{1}\frac{u^{4n+2}}{f(u)^{n+1}}~e^{(2n+1)(1-u^{2})\left(\frac{r_c}{r_*}\right)^{2}}du,\label{vt2}\\
V_{q\bar{q}}(r_*)&=\frac{r_*}{\pi \alpha'}\sum_{n=0}^{\infty}\frac{\Gamma(n+\frac{1}{2})}{\sqrt{\pi}\Gamma(n+1)}f(r_*)^{n} \int_{\delta}^{1}\frac{u^{4n-2}}{f(u)^{n}}~e^{(2n+(1-2n)u^{2})\left(\frac{r_c}{r_*}\right)^{2}} du.
\label{vt33}
\end{align}
Again, in order to find $V_{q\bar{q}}$ in terms of $l$ we should solve equation \eqref{vt2} for $r_*$ and use it in \eqref{vt33} to obtain $V_{q\bar{q}}$ as a function of $l$. Practically, it is not possible to solve equation \eqref{vt2} analytically to find $r_*$ as a function of $l$. However, we can study the behavior of meson potential energy in low and high temperature limits, in the high energy limit defined as $l\Lambda_{c}\ll 1$ (or $r_c \ll r_*$).
\subsubsection{Low temperature limit}\label{ltv}
At low temperature ($lT\ll 1$), the extremized classical string worldsheet is positioned next to the boundary and thus the leading contribution comes from the near boundary expansion. Finite temperature corrections and non-conformal effects appear as sub-leading terms corresponding to the deviation of the bulk geometry from pure AdS. Focusing on high energy limit, and using equation \eqref{vt2}, we reach the following expression for $l$ up to the 4th order in $r_c/r_{*}$ and $r_h/r_{*}$  
\begin{align}
l(r_{*})&=\frac{2}{r_*}\bigg[a_1+\alpha_{1}\bigg(\frac{r_h}{r_*}\bigg)^{4}+\bigg(a_2+\alpha_2 \bigg(\frac{r_h}{r_*}\bigg)^{4}\bigg)\bigg(\frac{r_c}{r_*}\bigg)^{2} \cr &+\bigg(a_3+\alpha_{3}\bigg(\frac{r_h}{r_*}\bigg)^{4}\bigg)\bigg(\frac{r_c}{r_*}\bigg)^{4} \bigg], \,\,\,\,\,\,\,\,\alpha_{1},\alpha_{2},\alpha_{3}<0,
\label{vtt4}
\end{align}
where $\alpha_{1}$, $\alpha_{2}$ and $\alpha_{3}$ are numerical coefficients reported in table \ref{listt1} in the appendix \ref{A}. Solving equation \eqref{vtt4} perturbatively for $r_*$ we obtain a perturbative expression for meson potential energy by considering the finite parts (the details of the calculations are written in appendix \ref{A})
\begin{align}
\hat{V}_{q\bar{q}}(l\Lambda_{c},lT)\equiv\frac{\alpha'V_{q\bar{q}}(l,l\Lambda_{c},lT)}{\Lambda_{c}^{\frac{1}{2}}T^{\frac{1}{2}}}&=\frac{1}{l\Lambda_{c}^{\frac{1}{2}}T^{\frac{1}{2}}}\bigg[\bar{b}_{1}+\bar{\beta}_{1}(lT)^{4}+\frac{1}{2}(\bar{b}_{2}+\bar{\beta}_{2}(lT)^{4})(l\Lambda_{c})^{2}\cr&+\frac{1}{4}(\bar{b}_{3} +\bar{\beta}_{3}(lT)^{4})(l\Lambda_{c})^{4}\bigg],\,\,\,\,\,\bar{\beta}_{1}<0,\,\,\,\, \bar{\beta}_{2}, \bar{\beta}_{3}>0,
\label{vtt8}
\end{align}
where $\hat{V}_{q\bar{q}}$ is the dimensionless meson potential energy at low temperature and $\bar{\beta}_1$, $\bar{\beta}_2$ and $\bar{\beta}_3$ are numerical coefficients given in appendix \ref{A} and reported in the table \ref{listt1}. Note that we make the meson potential energy dimensionless in such a way that we can get the intuitive results by taking the limits $l\Lambda_{c}\rightarrow 0$ and/or $lT\rightarrow 0$ of \eqref{vtt8}. The first two terms indicate the known meson potential energy for pure AdS and AdS black hole, respectively, and the next two terms are thermal and non-conformal corrections. In the high energy and low temperature limit, if we fix $lT$ and increase $l\Lambda_{c}$ we will then find that $\hat{V}_{q\bar{q}}$ becomes less bounded and the same results will be achieved by fixing $l\Lambda_{c}$ and increasing $lT$. We summarize the results corresponding to the mentioned limits.
\begin{itemize} 
\item $l\Lambda_{c}\rightarrow 0$, $lT$ finite and small enough: We get the zero temperature meson potential energy as the leading term and the sub-leading term, the term including $\bar{\beta}_1<0$, is the finite temperature correction which makes the meson potential energy less bounded.
\item $lT\rightarrow 0$, $l\Lambda_{c}$ finite and small enough: The leading term is the conformal meson potential energy and the non-conformal terms, including $\bar{b}_2>0$ and $\bar{b}_3>0$ terms, appear as the sub-leading terms. Clearly, these corrections causes the meson potential energy to be less bounded.
\item $l\Lambda_{c}\rightarrow 0$ and $lT\rightarrow 0$ : Obviously, we face the conformal meson potential energy.
\end{itemize}
According to the gauge/gravity dictionary, there is a phase transition point where $r_h\rightarrow r_{c}$ \cite{Andreev2} which is equivalent to the limit $T\rightarrow \Lambda_{c}/\pi \sqrt{2}$ and we call it the transition limit. We would like to study the meson potential energy near this point, hence we take the transition limit of $\hat{V}_{q\bar{q}}$. Doing some algebra, up to 4th order in $l\Lambda_{c}$ we reach the following expression
\begin{align}
\hat{V}_{q\bar{q}}(l\Lambda_{c},lT)\bigg|_{T\rightarrow \frac{\Lambda_{c}}{\pi \sqrt{2}}}=\frac{1}{l\Lambda_{c}}\bigg[\bar{b}_{1}+\frac{\bar{b}_{2}}{2}(l\Lambda_{c})^{2}+\bigg (\frac{\bar{b}_{3}}{4}+\frac{\bar{\beta}_{1}}{4\pi^4}\bigg ) (l\Lambda_{c})^{4}\bigg].
\label{vvt9}
\end{align}
To get a better understanding of the meson potential energy near transition point, one can fix $l\Lambda_{c}$ and compare zero temperature potential energy $\tilde{V}_{q\bar{q}}$ with finite temperature one $\hat{V}_{q\bar{q}}$ in the transition limit. To do so, we should focus on the difference between equations \eqref{vv12} and \eqref{vvt9} 
\begin{align}
\bigg|\hat{V}_{q\bar{q}}(l\Lambda_{c},lT)\bigg|_{T\rightarrow\frac{\Lambda_{c}}{\pi \sqrt{2}}}-\bigg|\tilde{V}_{q\bar{q}}(l\Lambda_{c})\bigg|=\frac{|\bar{\beta}_{1}|}{4\pi^4}(l\Lambda_{c})^{3}>0.
\label{vt10}
\end{align}
Near the phase transition point, it is seen that the zero temperature dimensionless meson potential energy is less bounded that is the amount of energy required to break down the meson to quark and anti-quark is less. We will return to this issue later in the paper.
\subsubsection{High temperature limit}\label{htv}
At high temperature, i.e. $lT\gg1$ (or $r_*\sim r_h$), the extremized classical string worldsheet tends to reach the horizon and therefore the leading contribution is related to the near horizon background. Since we work in the high energy limit, $r_c\ll r_*$, the high temperature limit can be identified by $r_h\rightarrow r_*$. As $r_*$ touches the horizon the classical string minimizes its energy by splitting into two vertical strings ending at the horizon. At high temperature, $r_h$ approaches $r_*$ and hence $f(r_*)=1-r_{h}^{4}/r_*^{4}\rightarrow0$, one can see from \eqref{vt33} that the meson potential energy is zero and there is no more information regarding the high temperature expansion of the meson potential energy.
\section{Holographic Subregion Complexity}\label{dd}
In this section we would like to study HSC using volume prescription written in \eqref{CV} for the MAdS and MBH backgrounds, equations \eqref{1} and \eqref{2}, respectively.
 \subsection{Holographic subregion complexity: Zero temperature expansion} \label{5.1}
We  consider a strip-like  boundary  region $A$ in the $\vec{x}$ direction at a constant time slice. This region can be parameterized as
\begin{eqnarray}
-\frac{l}{2}\leq x\equiv x(r)\leq\frac{l}{2},\qquad -\frac{L}{2}\leq y,z\leq\frac{L}{2},\qquad \qquad L\gg l.
\label{entangling region}
\end{eqnarray} 
Extremal surface is translationally invariant along $y$ and $z$ axes and the profile of the surface on the bulk is $x(r)$.
If one identifies the quark anti-quark separation, characteristic length $l$, with subregion length, HSC of $A$ can be considered as the complexity of the probe meson in non-conformal vacuum. According to \eqref{CV}, we need to obtain the volume enclosed by subregion $A$ and RT surface which we call $V_{\gamma_{A}}$. At zero temperature using MAdS background \eqref{1} we get
\begin{equation}
V_{\gamma_{A}}=2L^{2}\int_{r_*}^{\infty}r^2 e^{\frac{c}{r^2}} x(r) dr,
\label{co1}
\end{equation}
where $L^{2}$ is the volume of corresponding $yz$ plane. To  find $x(r)$ we need to compute the area of the RT surface  
\begin{equation}
A=2L^{2}\int_{0}^{\frac{l}{2}}r^2 e^{\frac{3c}{4r^2}} \sqrt{r^2+\frac{r'^2}{r^2}} dx.
\label{co2}
\end{equation}
Solving equation of motion for $x(r)$, the expressions for $V_{\gamma_{A}}$ and $l$ are given by 
\begin{align}
l(r_*)&=\frac{2}{r_*}\int_{0}^{1}u^3e^{\frac{3}{2}\left(\frac{r_c}{r_*}\right)^{2}(1-u^2)} \left(1-e^{3\left(\frac{rc}{r_*}\right)^{2}(1-u^2)}u^{6}\right)^{-\frac{1}{2}}du, \label{coo4}\\  V_{\gamma_{A}}(r_*)&=2L^{2}r_*^2\int_{\delta}^{1}u^{-4}e^{2\left(\frac{rc}{r_*}\right)^{2}u^2}\int_{u}^{1}u^3e^{\frac{3}{2}\left(\frac{r_c}{r_*}\right)^{2}(1-u^2)} \left(1-e^{3\left(\frac{rc}{r_*}\right)^{2}(1-u^2)}u^{6}\right)^{-\frac{1}{2}}du,
\label{co4}
\end{align}
where again $u=r_*/r$ and $\delta$ is an ultraviolet cutoff introduced since the volume integral is divergent. This integration is not solvable analytically. Hence, we use the same method, systematic expansion, in section \ref{cc1} and then \eqref{coo4} and \eqref{co4} can be written as follows  
\begin{align}
l(r_{*})&=\frac{2}{r_*}\sum_{n=0}^{\infty}\frac{\Gamma(n+\frac{1}{2})}{\sqrt{\pi}\Gamma(n+1)} \int_{0}^{1} u^{6n+3}~e^{\frac{3}{2}(2n+1)(1-u^{2})\left(\frac{r_c}{r_*}\right)^{2}}  du,\label{coo6}\\   V_{\gamma_{A}}(r_*)&=2L^{2}r_{*}^{2}\sum_{n=0}^{\infty}\frac{\Gamma(n+\frac{1}{2})}{\sqrt{\pi}\Gamma(n+1)}\int_{\delta}^{1}u^{-4}e^{2\left(\frac{rc}{r_*}\right)^{2}u^2}\int_{u}^{1}u^{6n+3}~e^{\frac{3}{2}(2n+1)(1-u^{2})\left(\frac{r_c}{r_*}\right)^{2}}du.
\label{co6}
\end{align}
Again to find HSC as a function of $l$ we need to solve equation \eqref{coo6} for $r_*$ and then put it in equation \eqref{co6} to obtain $V_{\gamma_{A}}$ in terms of $l$. HSC can be computed using equation \eqref{CV}. In practice, this procedure can not be performed analytically and hence we focus on high energy limit, $r_c\ll r_*$.
\subsubsection{High energy limit}\label{hec}
In high energy limit, from equation \eqref{coo6} and keeping up to the 4th order in $r_c/r_*$ we get
\begin{align}
l(r_*)=\frac{2}{r_*}\bigg[k_1+k_2\left(\frac{r_c}{r_*}\right)^{2}+k_3\left(\frac{r_c}{r_*}\right)^{4}  \bigg],\,\,\,\,k_1,k_2,k_3>0,
\label{cco8}
\end{align}
where $k_1$, $k_2$ and $k_3$ are numerical coefficients reported in table \ref{list2} in appendix \ref{B}. If we solve equation \eqref{cco8} perturbatively for $r_*$ then we will finally obtain the following finite expression (for more details of the calculation refer to appendix \ref{B})
 \begin{equation}
\tilde{V}_{\gamma_{A}}(l\Lambda_{c})\equiv\frac{V_{\gamma_{A}}(l,l\Lambda_{c})}{L^2\Lambda_{c}^{2}}=\frac{1}{(l\Lambda_{c})^{2}}\bigg(\bar{w}_{1}+\frac{\bar{w}_{2}}{2}(l\Lambda_{c})^{2}+\frac{\bar{w}_{3}}{4}(l\Lambda_{c})^{4}  \bigg),\,\,\,\,\,\bar{w}_{1}, \bar{w}_{2}, \bar{w}_{3}<0,
\label{cco12}
\end{equation}
where $\tilde{V}_{\gamma_{A}}$ is dimensionless volume and $\bar{w}_{1}$, $\bar{w}_{2}$ and $\bar{w}_{3}$ are numerical coefficients shown in table \ref{list2} in appendix \ref{B}. Finally, HSC can be obtained using equations \eqref{CV} and \eqref{cco12}. The first negative term contributes to the boundary of MAdS. The second and third negative terms correspond to the non-conformality effects. These negative non-conformal corrections which is added to the first negative term, increase the absolute value of the dimensionless volume which means the non-conformal state needs less information to be specified.
In other words, the more non-conformality effects, the less required information to specify the meson state. On the other hand, as we mentioned in the subsection \ref{hev}, the more non-conformality effects make less bounded dimensionless meson potential energy.

\subsection{Holographic subregion complexity: Finite temperature expansion}\label{ct}
Here, we investigate the thermal behavior of HSC in the MBH background \eqref{2}. We develop a systematic expansion in low temperature $lT\ll 1$ (or $r_h\ll r_*$) and high temperature $lT\gg 1$ (or $r_* \sim r_{h}$) limits and focus on the high energy limit $l\Lambda_{c}\ll 1$ (or $r_{c}\ll r_{*}$). The same calculations lead to the following expressions for $l$ and $V_{\gamma_{A}}$, 
\begin{align}
l(r_*)&=\frac{2}{r_*}\int_{0}^{1}\frac{u^{3}}{\sqrt{f(u)}}e^{\frac{3}{2}\left(\frac{r_c}{r_*}\right)^{2}(1-u^2)} \left(1-e^{3\left(\frac{rc}{r_*}\right)^{2}(1-u^2)}u^{6}\right)^{-\frac{1}{2}}du,\label{cott1}\\ 
V_{\gamma_{A}}(r_*)&=2L^{2}r_*^2\int_{\delta}^{1} \frac{u^{-4}}{\sqrt{f(u)}}e^{2\left(\frac{rc}{r_*}\right)^{2}u^2}\int_{u}^{1}\frac{u^{3}}{\sqrt{f(u)}} e^{\frac{3}{2}\left(\frac{r_c}{r_*}\right)^{2}(1-u^2)} \left(1-e^{3\left(\frac{r_c}{r_*}\right)^{2}(1-u^2)}u^{6}\right)^{-\frac{1}{2}}du,
\label{cot1}
\end{align}
where $\delta$ is an ultraviolet cutoff and $f(u)=1-(r_h/r_*)^{4}u^{4}$. Again, analogous to the section \ref{fte}, the above integrals can not be solved analytically and hence we can develop a systematic expansion. Using binomial expansions equations \eqref{cott1} and \eqref{cot1} are given by
\begin{align}
l(r_*)&=\frac{2}{r_*}\sum_{n=0}^{\infty}\sum_{m=0}^{\infty}\frac{\Gamma(n+\frac{1}{2})\Gamma(m+\frac{1}{2})}{\pi\Gamma(n+1)\Gamma(m+1)}\int_{0}^{1}\bigg(\frac{r_h}{r_*}\bigg)^{4m}u^{4m+6n+3}e^{\frac{3}{2}(2n+1)(1-u^{2})\left(\frac{r_c}{r_*}\right)^{2}}du,\label{cot22}\\
V_{\gamma_{A}}(r_*)&=2V_{y,z}r_*^2\sum_{n=0}^{\infty}\sum_{m=0}^{\infty}\sum_{p=0}^{\infty}\frac{\Gamma(n+\frac{1}{2})\Gamma(m+\frac{1}{2})\Gamma(p+\frac{1}{2})}{\pi^{3/2}\Gamma(n+1)\Gamma(m+1)\Gamma(p+1)}\int_{\delta}^{1}\left(\frac{r_{h}}{r_{*}}\right)^{4p}u^{4p-4}e^{2\left(\frac{r_c}{r_*}\right)^{2}u^2} \cr & \times 
\int_{u}^{1}\left(\frac{r_h}{r_*}\right)^{4m}u^{4m+6n+3}e^{\frac{3}{2}(2n+1)(1-u^{2})\left(\frac{r_c}{r_*}\right)^{2}}du.\label{cot33}
\end{align}
Since $r_{c}<r_{*}$ and $r_{h}<r_{*}$ we do not worry about the convergence of the above series and hence these are well-defined. Now, the rest of the procedure is familiar. We have to take the following steps: Solving equation \eqref{cot22} for $r_*$ and then calculating $V_{\gamma_{A}}$ by using equation \eqref{cot33} and finally computing HSC using equation \eqref{CV}. Unfortunately, this process can not be done analytically and hence we have to study the behaviour of the HSC at low and high temperature, in the high energy limit.
\subsubsection{low temperature limit}\label{ltc}
\begin{figure}[tbp]
\centering
\includegraphics[width=75 mm]{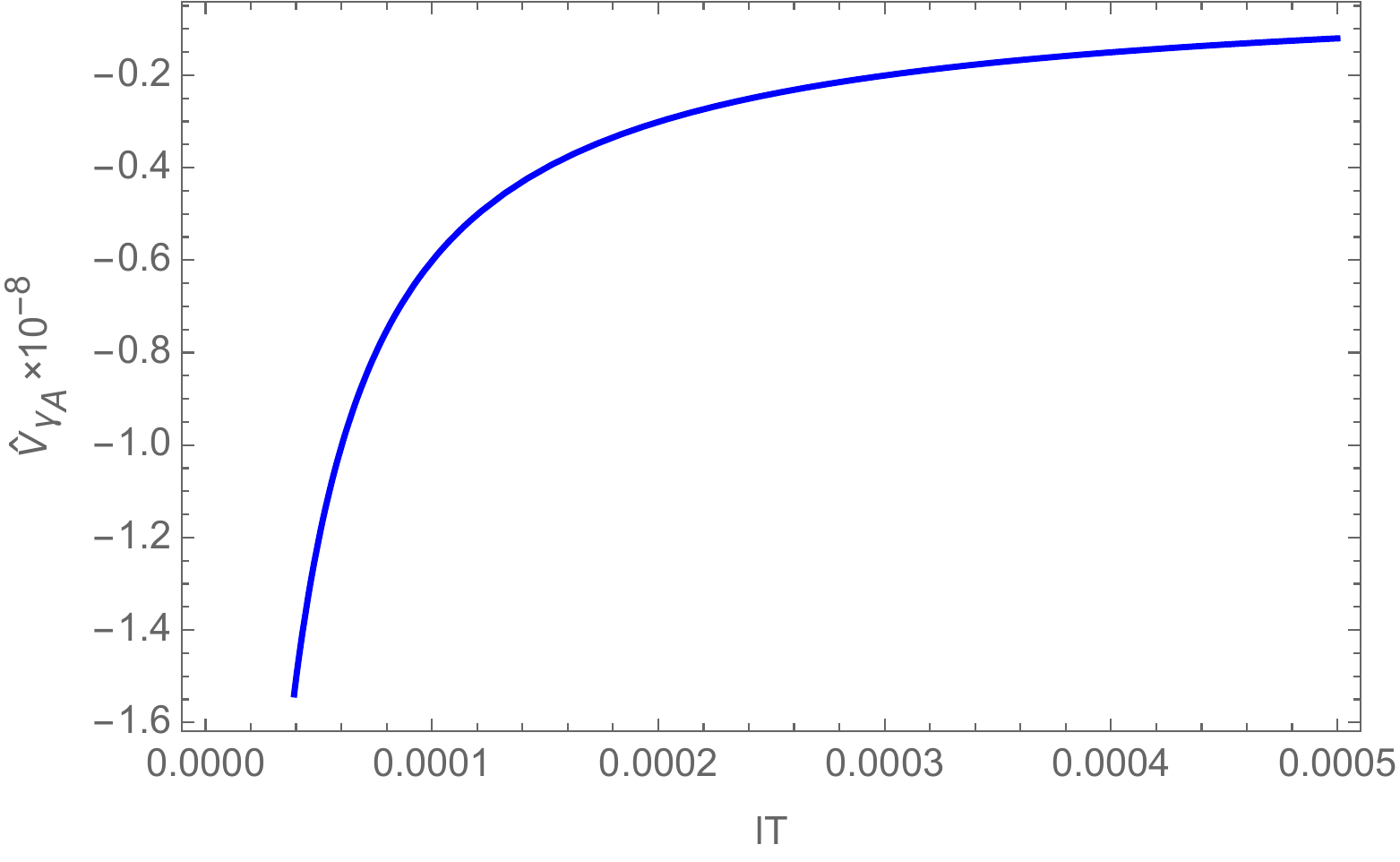}
\includegraphics[width=75 mm]{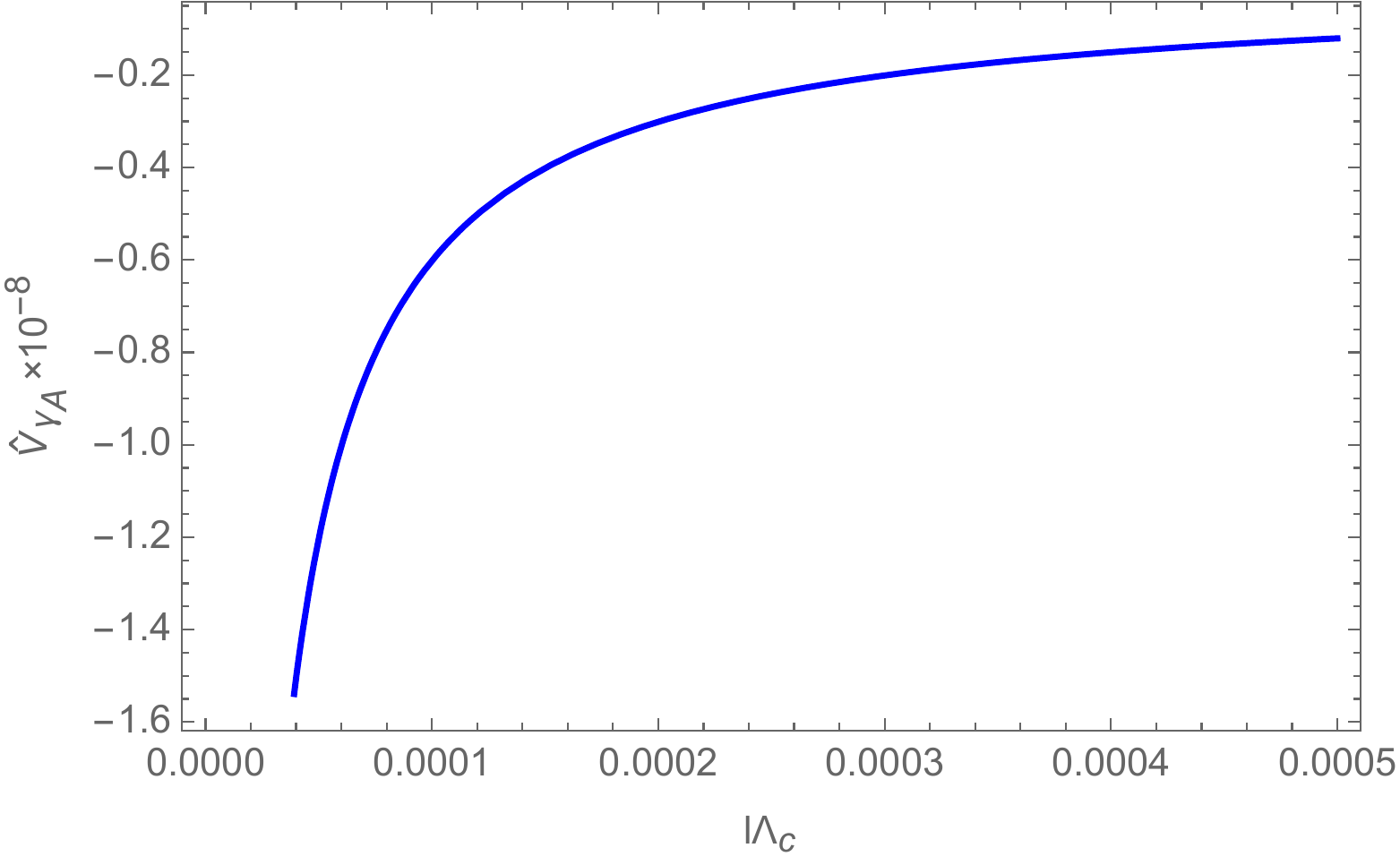}
\caption{Left: At leading order, the dimensionless volume $\hat{V}_{\gamma_{A}}$ as a function of $lT$ for fixed $l\Lambda_{c}=0.0001$, in the low temperature limit. Right: The dimensionless subregion volume $\hat{V}_{\gamma_{A}}$ as a function of $l\Lambda_{c}$ for fixed $lT=0.0001$, in the low temperature limit.}
\label{fig2}
\end{figure}

At low temperature $lT\ll 1$, the extremal surface is restricted to be near the boundary and hence the leading contribution to the HSC comes from near AdS boundary expansion. Therefore, we should expect the zero temperature HSC to be the leading term. Finite temperature corrections correspond to the deviation of the bulk geometry from pure AdS. Working on the high energy limit, and using \eqref{cot22}, we get the following expression for $l$ up to the 4th order in $r_c/r_*$ and $r_h/r_*$ we have 
\begin{align}
l(r_*)&=\frac{2}{r_*}\bigg[k_1+\kappa_{1}\bigg(\frac{r_h}{r_*}\bigg)^{4}+\bigg(k_2+\kappa_{2}\bigg(\frac{r_h}{r_*}\bigg)^{4}\bigg)\bigg(\frac{r_c}{r_*}\bigg)^{2}+\cr &\bigg(k_3+\kappa_{3}\bigg(\frac{r_h}{r_*}\bigg)^{4}\bigg)\bigg(\frac{r_c}{r_*}\bigg)^{4} \bigg],\,\,\,\,\,\,\kappa_{1},\kappa_{2},\kappa_{3}>0
\label{ccot4}
\end{align}

where $\kappa_{1},\kappa_{2},\kappa_{3}$ are numerical coefficients reported in table \ref{listt2} in appendix \ref{B}. Solving equation \eqref{ccot4} perturbatively for $r_*$ we reach a perturbative finite expression
\begin{align}
\hat{V}_{\gamma_{A}}(l\Lambda_{c},lT)\equiv\frac{V_{\gamma_{A}}(l,l\Lambda_{c},lT)}{L^{2}\Lambda_{c}T}&=\frac{1}{l^{2}\Lambda_{c}T}\bigg(\bar{w}_{1}+\bar{\omega}_{1}(lT)^{4}+\frac{1}{2}(\bar{w}_2+\bar{\omega}_{2}(lT)^{4})(l\Lambda_{c})^{2}\cr &+\frac{1}{4}(\bar{w}_3+\bar{\omega}_{3}(lT)^{4})(l\Lambda_{c})^{4}\bigg),\,\,\,\,\bar{\omega}_{1},\bar{\omega}_{2}<0\,\,\,,\bar{\omega}_{3}>0,
\label{ccot8}
\end{align}
where $\hat{V}_{\gamma_{A}}$ is dimensionless volume at low temperature and $\bar{\omega}_{1},\bar{\omega}_{2}$ and $\bar{\omega}_{3}$ are numerical coefficients reported in table \ref{listt2} in appendix \ref{B}, the details of the calculation are in appendix \ref{B}. HSC can be computed using the relation \eqref{CV}. We make $V_{\gamma_{A}}$ dimensionless to have meaningful and intuitive limits which are favorable such as $l\Lambda_{c}\rightarrow 0$ and/or $lT\rightarrow 0$. The first two terms are the known results corresponding to the pure AdS and AdS black hole HSC, respectively, and the next two terms indicate the thermal and non-conformal corrections. From equation \eqref{ccot8}, it is seen that if one fixes $lT$ and increases $l\Lambda_{c}$, then $\hat{V}_{\gamma_{A}}$ will decrease and the same results will be obtained by changing the role of $lT$ and $l\Lambda_{c}$. This result can be clearly observed from figure \ref{fig2} where we plot $\hat{V}_{\gamma_{A}}$ as a function of $lT$ ($l\Lambda_{c}$) for fixed value of $l\Lambda_{c}$ ($lT$). Interestingly, following the results we get for $\hat{V}_{q\bar{q}}$ in the subsection \ref{ltv}, one can conclude that the less bounded $\hat{V}_{q\bar{q}}$ the less information required to specify the meson state, at leading order, and the thermal effects may not respect to this result. Below, we represent the results corresponding to the desired limits of equation \eqref{ccot8}.
\begin{itemize} 
\item $l\Lambda_{c}\rightarrow 0$, $lT$ finite and small enough: The leading term corresponding to the zero temperature HSC and the sub-leading term, the term including $\bar{\omega}_{1}<0$, is the finite temperature correction which leads to a decline in corresponding HSC.
\item $lT\rightarrow 0$, $l\Lambda_{c}$ finite and small enough: The leading term corresponding to the conformal HSC and the non-conformal corrections, including $\bar{w}_2<0$ and $\bar{w}_3<0$ terms, appear as the sub-leading terms. These corrections lead to a decrease in corresponding HSC.
\item $l\Lambda_{c}\rightarrow 0$ and $lT\rightarrow 0$: Clearly, we reach the corresponding conformal HSC.
\end{itemize}

We would like to study HSC near the transition point, by taking the transition limit $T\rightarrow \Lambda_{c}/\pi \sqrt{2}$ of equation \eqref{ccot8}. Up to the fourth order in $l\Lambda_{c}$ we get
\begin{align}
\hat{V}_{\gamma_{A}}=(l\Lambda_{c},lT)\bigg|_{T\rightarrow\frac{\Lambda_{c}}{\pi\sqrt{2}}}=\frac{1}{(l\Lambda_{c})^{2}}\bigg[\bar{w}_{1}+\frac{\bar{w}_2}{2}(l\Lambda_{c})^{2}+\bigg(\frac{\bar{w}_{3}}{4}+\frac{\bar{\omega}_{1}}{4\pi^{4}}\bigg)(l\Lambda_{c})^{4} \bigg].
\label{cot11}
\end{align}
We fix $l\Lambda_{c}$ and compare the HSC at zero temperature \eqref{cco12} and finite temperature \eqref{cot11} in the transition limit. We have
\begin{align}
\bigg|\hat{V}_{\gamma_{A}}(l\Lambda_{c},lT)\bigg|_{T\rightarrow \frac{\Lambda_{c}}{\pi \sqrt{2}}}-\bigg|\tilde{V}_{\gamma_{A}}(l\Lambda_{c})\bigg| =\frac{|\bar{\omega}_{1}|}{4\pi^4}(l\Lambda_{c})^{3}>0.
\label{cot12}
\end{align}
If we accept that a state which needs less information to be specified is a \textit{favorable state}, then equation \eqref{cot12} will indicate that near the phase transition point the meson state at zero temperature is favorable. Additionally, according to the equation \eqref{vt10} the zero temperature meson state is less bounded. Therefore, in short, near the transition point the less bounded meson is more favorable, i.e. it needs less information to specify in the mentioned limits. Notice that by taking the transition limit, the meson state at finite temperature is more bounded. These results is in agreement with \cite{sub:2019}.

\subsubsection{High temperature limit}\label{htc}
At high temperature, i.e. $lT\gg 1$ (or ${r_*}\sim{r_h}$), the thermal fluctuations become considerable and the extremal surface gets close to the horizon. The leading contribution comes from the near horizon background and the full bulk contributes to the subleading terms. In high temperature and high energy limit $r_c\ll r_*$ and $r_h\rightarrow r_{*}$, respectively, we integrate equations \eqref{cot22} and \eqref{cot33} and consider finite parts, up to $(r_c/r_*)^2$ we get
\begin{align}
l(r_*)=\frac{2}{r_*}\sum_{n=0}^{\infty}\sum_{m=0}^{\infty}\frac{\Gamma(n+\frac{1}{2})\Gamma(m+\frac{1}{2})}{\pi\Gamma(n+1)\Gamma(m+1)}\bigg(\frac{r_h}{r_*}\bigg)^{4m}\bigg[L_{1}+L_{2}\bigg(\frac{r_c}{r_*}\bigg)^{2}\bigg],
\label{h1}
\end{align}
\begin{align}
V_{\gamma_{A}}&=2L^{2}r_{*}^{2}\sum_{n=0}^{\infty}\sum_{m=0}^{\infty}\sum_{p=0}^{\infty}\frac{\Gamma(n+\frac{1}{2})\Gamma(m+\frac{1}{2})\Gamma(p+\frac{1}{2})}{\pi^{3/2}\Gamma(n+1)\Gamma(m+1)\Gamma(p+1)}\bigg(\frac{r_h}{r_*}\bigg)^{4(m+p)} \bigg[C_{1}+C_{2}\bigg(\frac{r_c}{r_*}\bigg)^{2}\bigg],
\label{h2}
\end{align}
where $L_{1}$, $L_{2}$, $C_{1}$ and $C_{2}$ are constant coefficients given by
\begin{align}
&L_{1}=\frac{1}{6n+4m+4}, \cr &
L_{2}=\frac{3}{2}(2n+1)\bigg(\frac{1}{6n+4m+4}-\frac{1}{6n+4m+6}\bigg),\cr &
C_{1}=\frac{1}{(6n+4m+4)(4p-3)}-\frac{1}{(6n+4m+4)(6n+4m+4p+1)}, \cr & 
C_{2}=\frac{3}{2}(2n+1)\bigg(\frac{1}{(6n+4m+4)(4p-3)}-\frac{1}{(6n+4m+4)(6n+4m+4p+1)} \cr&~~~~  -\frac{1}{(6n+4m+6)(4p-3)}+\frac{1}{(6n+4m+6)(6n+4m+4p+3)}\bigg)\cr &~~~~ +\frac{2}{(6n+4m+4)(4p-1)}-\frac{2}{(6n+4m+4)(6n+4m+4p+3)}.
\label{h4}
\end{align}
The next step is to write $V_{\gamma_{A}}$ in terms of $l$. To do so, we perform the calculations order by order up to $(r_c/r_*)^2$ and check carefully the convergence of the resulting infinite series.
\begin{itemize}
\item{Up to $(r_c/r_*)^{0}$:} We consider the terms including $L_{1}$ and $C_{1}$ which are the known results for AdS black hole. Using equation \eqref{h1} and \eqref{h2}, we obtain
\begin{align}
\frac{V_{\gamma_{A}}}{L^{2}}=lr_{*}^{3}\bigg(\sum_{p=1}^{\infty}\frac{\Gamma(p+\frac{1}{2})}{\sqrt{\pi}\Gamma(p+1)(4p-3)}\bigg(\frac{r_h}{r_*}\bigg)^{4p}-\frac{1}{3}\bigg)-2r_{*}^{2}\cal{A},
\label{h5}
\end{align}
where the term $-1/3$ corresponds to $p=0$ term and expresses the fact that only $p=0$ term is divergent and ${\cal{A}}$ is given by
\begin{align}
{\cal{A}}=\sum_{n=0}^{\infty}\sum_{m=0}^{\infty}\sum_{p=0}^{\infty}\frac{\Gamma(n+\frac{1}{2})\Gamma(m+\frac{1}{2})\Gamma(p+\frac{1}{2})}{\pi^{3/2}\Gamma(n+1)\Gamma(m+1)\Gamma(p+1)(6n+4m+4)(6n+4m+4p+1)}\bigg(\frac{r_h}{r_*}\bigg)^{4(m+p)}.
\label{h6}
\end{align}
For large $p$, the first infinite series in \eqref{h5} goes as $\sim$ $p^{-3/2}(r_h/r_*)^{4p}$ and hence converges for $r_h=r_*$. We are in a position to check the convergence of \eqref{h6}, then we sum over $n$ and $p$ in \eqref{h6} and obtain
\begin{align}
{\cal{A}}=&\sum_{m=0}^{\infty}\bigg(\frac{r_h}{r_*}\bigg)^{4m}\frac{\Gamma(m+\frac{1}{2})}{720\Gamma(m+1)}\bigg[-\frac{40\Gamma(\frac{2m}{3}+\frac{2}{3})}{\Gamma(\frac{2m}{3}+\frac{7}{6})}{_{2}F_{1}}\bigg(-\frac{3}{4},\frac{1}{2};\frac{1}{4};\bigg(\frac{r_h}{r_*}\bigg)^{4}\bigg)+\frac{40\Gamma(\frac{2m}{3}+\frac{1}{6})}{\Gamma(\frac{2m}{3}+\frac{2}{3})}\cr &{_{6}F_{5}}\bigg(-\frac{1}{4},\frac{1}{6},\frac{1}{2},\frac{5}{6},\frac{m}{3}+\frac{1}{12},\frac{m}{3}+\frac{7}{12};\frac{1}{3},\frac{2}{3},\frac{3}{4},\frac{m}{3}+\frac{1}{3},\frac{m}{3}+\frac{5}{6};\bigg(\frac{r_h}{r_*}\bigg)^{12}\bigg)-\frac{60\Gamma(\frac{2m}{3}+\frac{5}{6})}{\Gamma(\frac{2m}{3}+\frac{4}{3})}\cr &\bigg(\frac{r_h}{r_*}\bigg)^{4}{_{6}F_{5}}\bigg(\frac{1}{12},\frac{1}{2},\frac{5}{6},\frac{7}{6},\frac{m}{3}+\frac{5}{12},\frac{m}{3}+\frac{11}{12};\frac{2}{3},\frac{13}{12},\frac{4}{3},\frac{m}{3}+\frac{2}{3},\frac{m}{3}+\frac{7}{6};\bigg(\frac{r_h}{r_*}\bigg)^{12}\bigg)-\frac{9\Gamma(\frac{2m}{3}+\frac{3}{2})}{\Gamma(\frac{2m}{3}+2)}\cr & \bigg(\frac{r_h}{r_*}\bigg)^{8}{_{6}F_{5}}\bigg(\frac{5}{12},\frac{5}{6},\frac{7}{6},\frac{3}{2},\frac{m}{3}+\frac{3}{4},\frac{m}{3}+\frac{5}{4};\frac{4}{3},\frac{17}{12},\frac{5}{3},\frac{m}{3}+1,\frac{m}{3}+\frac{3}{2};\bigg(\frac{r_h}{r_*}\bigg)^{12}\bigg)\bigg],
\label{h7}
\end{align}
where $_{p}F_{q}$ is the hypergeometric function and we use the following relation, for $q=p-1$
\begin{align}
&_{p}F_{q}\bigg(a_{1},...,a_{p};b_{1},...,b_{q};z\bigg)=\sum_{k=0}^{\infty}\frac{z^{k}}{k!}\frac{(a_{1})_{k}...(a_{p})_{k}}{(b_{1})_{k}...(b_{q})_{k}}, \,\,\,\, |z|<1 \cr &~~~~~~~~~~  {\rm{or}} \,\,\,\, |z|=1 \,\,\ {\rm{and}} \,\,\,\,\, {\rm{Re}}\bigg(\sum_{j=1}^{q}b_{j}-\sum_{j=1}^{p}a_{j}\bigg)>0,
\label{h8}
\end{align}
where $(a)_{k}\equiv\Gamma(a+k)/\Gamma(a)$ is the Pochhamer symbol. Using equation \eqref{h8} it can be shown that for large $m$, each term of the sum \eqref{h7} behaves as $\sim m^{-1}(r_{h}/r_*)^{4m}$ and hence the series diverges for $r_*=r_h$. We should add and subtract the divergence piece of this series and then write $r_*=r_h(1+\epsilon)$ where at high temperature $\epsilon\ll 1$ and do an expansion for small $\epsilon$. We obtain
\begin{align}
{\cal{A}}=A_{1}+A_{2}~{\rm{ln}}(4\epsilon)+{\cal{O}}(\epsilon),
\label{h9}
\end{align}
where $A_{1}$ and $A_{2}$ are constant coefficients given in appendix \ref{B} and ${\rm{ln}}(4\epsilon)$ is a function of $r_h$ and $l$ obtained in appendix \ref{B}. Using \eqref{h5} and \eqref{h9}, we reach the following linear relation
\begin{align}
V_{HT}^{(0)}(lT)\equiv\frac{V_{\gamma_{A}}}{L^{2}T^{2}}=M_{1}^{(0)}+lTM_{2}^{(0)},
\label{h10}
\end{align}
where $V_{HT}^{(0)}$ is dimensionless volume up to the $(r_c/r_*)^{0}$ at high temperature, $M_{1}^{(0)}$ and $M_{2}^{(0)}$ are
constant coefficients given by
\begin{align}
&M_{1}^{(0)}=-2\pi^{2}\bigg[A_{1}+A_{2}\bigg(\frac{\sqrt{6\pi}\Gamma(\frac{2}{3})}{3\Gamma(\frac{7}{6})}+\sum_{m=1}^{\infty}\bigg(\frac{\sqrt{6}\Gamma(m+\frac{1}{2})\Gamma(\frac{2m}{3}+\frac{2}{3})}{3\Gamma(m+1)\Gamma(\frac{2m}{3}+\frac{7}{6})}-\frac{1}{m}\bigg)\bigg)\bigg],\cr &M_{2}^{(0)}=\sum_{p=1}^{\infty}\frac{\pi^{5/2}\Gamma(p+\frac{1}{2})}{\Gamma(p+1)(4p-3)}-\frac{\pi^{3}}{3}+2\sqrt{6}\pi^{3}A_{2}. 
\label{h11}
\end{align}
Finally, HSC can be obtained using equations \eqref{CV} and \eqref{h10}. According to equation \eqref{h10}, at high temperature, the term including $M_{2}^{(0)}$ is dominant indicating $V_{HT}^{(0)}$ is proportional to $lT$ and the term independent of $lT$ receives contribution from the full bulk background.
\item{Up to $(r_{c}/r_*)^{2}$:} We consider the terms including $L_{2}$ and $C_{2}$ which are the non-conformal effects at high temperature. Using \eqref{h1} and \eqref{h2}, we have
\begin{align}
\frac{V_{\gamma_{A}}}{L^{2}}=&lr_{*}^{3}\bigg(\sum_{p=1}^{\infty}\frac{\Gamma(p+\frac{1}{2})}{\sqrt{\pi}\Gamma(p+1)(4p-3)}\bigg(\frac{r_h}{r_*}\bigg)^{4p}-\frac{1}{3}\bigg)+2r_{*}^{2}\bigg({\cal{B}}\bigg(\frac{r_c}{r_*}\bigg)^{2}-{\cal{A}}\bigg),
\label{h13}
\end{align}
where again the term $-1/3$ comes from $p=0$ term which is the divergent part and ${\cal{B}}$ is given by
\begin{align}
&{\cal{B}}=\sum_{n=0}^{\infty}\sum_{m=0}^{\infty}\sum_{p=0}^{\infty}\frac{\Gamma(n+\frac{1}{2})\Gamma(m+\frac{1}{2})\Gamma(p+\frac{1}{2})}{\pi^{3/2}\Gamma(n+1)\Gamma(m+1)\Gamma(p+1)}\bigg[\frac{3}{2}(2n+1)\cr &~~~~~~~~~~~~~~~~~~~~~~\times\bigg(\frac{1}{(6n+4m+6)(6n+4m+4p+3)}\cr &~~~~~~~~~~~~~~~~~~~~~~-\frac{1}{(6n+4m+4)(6n+4m+4p+1)}\bigg)\cr &~~~~~~~~~~~~~~~~~~~~~~+\frac{2}{(6n+4m+4)(4p-1)}\cr &~~~~~~~~~~~~~~~~~~~~~~-\frac{2}{(6n+4m+4)(6n+4m+4p+3)}   \bigg]\bigg(\frac{r_h}{r_*}\bigg)^{4(m+p)}.
\label{h14}
\end{align}
In order to study the convergence of the series in \eqref{h14} at high temperature, $r_*\sim r_h$, we divide ${\cal{B}}$ into the two infinite series called ${\cal{B}}_{1}$ and ${\cal{B}}_{2}$ which is investigated respectively. First we study the behaviour of ${\cal{B}}_{1}$ whose form is written by 
\begin{align}
{\cal{B}}_{1}&=\sum_{n=0}^{\infty}\sum_{m=0}^{\infty}\sum_{p=0}^{\infty}\frac{\Gamma(n+\frac{1}{2})\Gamma(m+\frac{1}{2})\Gamma(p+\frac{1}{2})}{\pi^{3/2}\Gamma(n+1)\Gamma(m+1)\Gamma(p+1)}\frac{3}{2}(2n+1)\cr &~~~~~~~~~~~~~~~~~~~~ \times\bigg(\frac{1}{(6n+4m+6)(6n+4m+4p+3)}\cr &~~~~~~~~~~~~~~~~~~~~ -\frac{1}{(6n+4m+4)(6n+4m+4p+1)}\bigg)\bigg(\frac{r_h}{r_*}\bigg)^{4(m+p)}\cr &=\sum_{n=0}^{\infty}\sum_{m=0}^{\infty}\bigg[\frac{3(2n+1)\Gamma(n+\frac{1}{2})\Gamma(m+\frac{1}{2})}{4\pi (3n+2m+3)(6n+4m+3)\Gamma(n+1)\Gamma(m+1)}\cr &~~~~~~~~~~~~~~~~ {_{2}F_{1}}\bigg(\frac{1}{2},m+\frac{3n}{2}+\frac{3}{4};m+\frac{3n}{2}+\frac{7}{4};\bigg(\frac{r_h}{r_*}\bigg)^{4}\bigg)\cr &~~~~~~~~~~~~~~~~   -\frac{3(2n+1)\Gamma(n+\frac{1}{2})\Gamma(m+\frac{1}{2})}{4\pi (3n+2m+2)(6n+4m+1)\Gamma(n+1)\Gamma(m+1)}\cr &~~~~~~~~~~~~~~~~ {_{2}F_{1}}\bigg(\frac{1}{2},m+\frac{3n}{2}+\frac{1}{4};m+\frac{3n}{2}+\frac{5}{4};\bigg(\frac{r_h}{r_*}\bigg)^{4}\bigg)\bigg]\bigg(\frac{r_h}{r_*}\bigg)^{4m},
\label{h15}
\end{align}
where in the last equation we sum over $p$. For large $n$ and $m$, there is an equal behavior regarding the two terms in \eqref{h15} due to their functional form and both of them converge at high temperature, $r_*\sim r_h$ . We are left to check the convergence of the remaining part of ${\cal{B}}$, called ${\cal{B}}_{2}$, which is given by the following expression
\begin{align}
{\cal{B}}_{2}&=\sum_{n=0}^{\infty}\sum_{m=0}^{\infty}\sum_{p=0}^{\infty}\frac{\Gamma(n+\frac{1}{2})\Gamma(m+\frac{1}{2})\Gamma(p+\frac{1}{2})}{\pi^{3/2}\Gamma(n+1)\Gamma(m+1)\Gamma(p+1)}\bigg(\frac{2}{(6n+4m+4)(4p-1)}\cr &~~~~~~~~~~~~~~~~~~~-\frac{2}{(6n+4m+4)(6n+4m+4p+3)}\bigg)\bigg(\frac{r_h}{r_*}\bigg)^{4(m+p)} \cr &=\sum_{m=0}^{\infty}\bigg(\frac{r_h}{r_*}\bigg)^{4m}\bigg[\frac{\Gamma(m+\frac{1}{2})\Gamma(\frac{2m}{3}+\frac{3}{2})}{84(4m+3)(4m+7)(4m+11)\Gamma(m+1)\Gamma(\frac{2m}{3}+1)} \cr &~~~~~~~~~~~{_{6}F_{5}}\bigg(-\frac{1}{12},\frac{1}{6},\frac{1}{2},\frac{5}{6},\frac{m}{3}+\frac{1}{4},\frac{m}{3}+\frac{3}{4};\frac{1}{3},\frac{2}{3},\frac{11}{12},\frac{m}{3}+\frac{1}{2},\frac{m}{3}+1;\bigg(\frac{r_h}{r_*}\bigg)^{12}\bigg)\cr &~~~~~~~~~~~(-12936-12096~m-2688~m^2)+\frac{\Gamma(m+\frac{1}{2})\Gamma(\frac{2m}{3}+\frac{13}{6})}{84(4m+3)(4m+7)(4m+11)\Gamma(m+1)\Gamma(\frac{2m}{3}+\frac{5}{3})}\cr &~~~~~~~~~~~~{_{6}F_{5}}\bigg(\frac{1}{4},\frac{1}{2},\frac{5}{6},\frac{7}{6},\frac{m}{3}+\frac{7}{12},\frac{m}{3}+\frac{13}{12};\frac{2}{3},\frac{5}{4},\frac{4}{3},\frac{m}{3}+\frac{5}{6},\frac{m}{3}+\frac{4}{3};\bigg(\frac{r_h}{r_*}\bigg)^{12}\bigg)\bigg(\frac{r_h}{r_*}\bigg)^{4}\cr&~~~~~~~~~~~(924+1568~m+448~m^2)+\frac{\Gamma(m+\frac{1}{2})\Gamma(\frac{2m}{3}+\frac{17}{6})}{84(4m+3)(4m+7)(4m+11)\Gamma(m+1)\Gamma(\frac{2m}{3}+\frac{7}{3})}\cr &~~~~~~~~~~~~{_{6}F_{5}}\bigg(\frac{7}{12},\frac{5}{6},\frac{7}{6},\frac{3}{2},\frac{m}{3}+\frac{11}{12},\frac{m}{3}+\frac{17}{12};\frac{4}{3},\frac{19}{12},\frac{5}{3},\frac{m}{3}+\frac{7}{6},\frac{m}{3}+\frac{5}{3};\bigg(\frac{r_h}{r_*}\bigg)^{12}\bigg)\bigg(\frac{r_h}{r_*}\bigg)^{8} \cr&~~~~~~~~~~~(189+360~m+144~m^2)\bigg],
\label{h16}
\end{align}
where in the last equation we sum over $n$ and $p$. There exist three different types of behavior for large $m$ in \eqref{h16}. There are specific terms whose large $m$ behavior goes as $\sim m^{-3}(r_h/r_*)^{4m}$ and $\sim m^{-2}(r_h/r_*)^{4m}$ which are clearly convergent at $r_*=r_h$, by isolation of the term corresponding to $m=0$ of the series. However, there are other terms which go as $\sim m^{-1}(r_h/r_*)^{4m}$ and hence they are divergent at $r_*=r_h$. In order to get convergent series we should extract the divergence piece. Similar to the previous calculation of the zero order of $r_c/r_*$, we finally get
\begin{align}
{\cal{B}}=B_{1}+B_{2}~{\rm{ln}}(4\epsilon)+{\cal{O}}(\epsilon),
\label{h17}
\end{align}
\begin{figure}[tbp]
\centering
\includegraphics[width=75 mm]{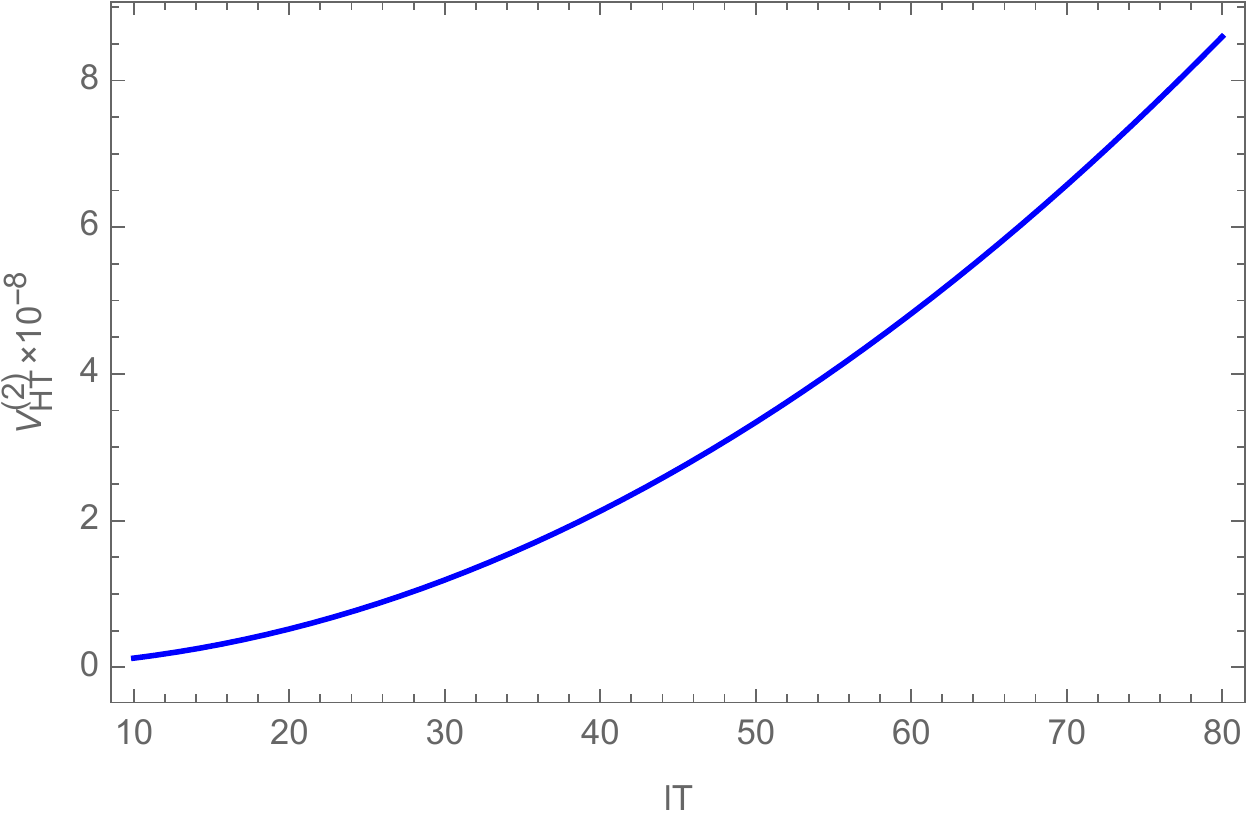} 
\includegraphics[width=75 mm]{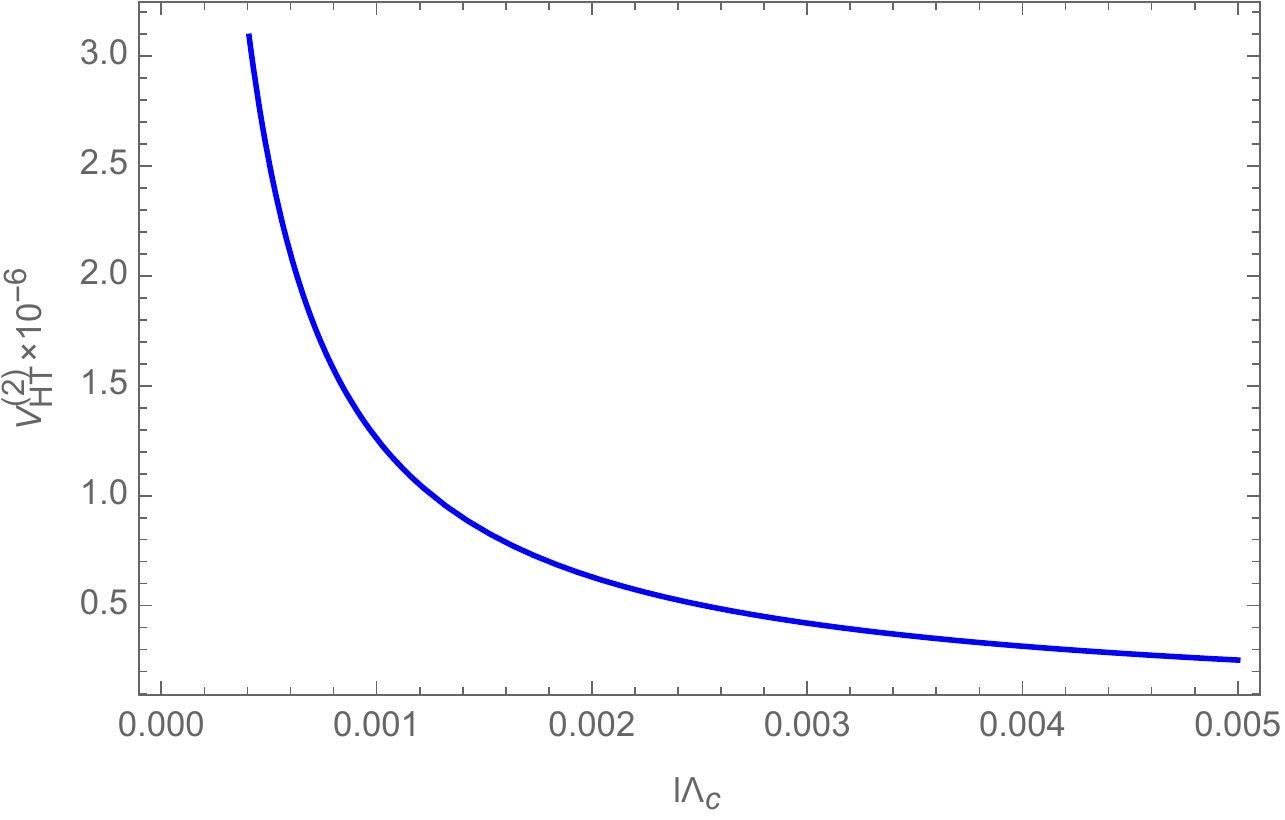}
\caption{Left: The dimensionless volume $V_{HT}^{(2)}$ as a function of $lT$ for fixed $l\Lambda_{c}=0.0001$, in the high temperature limit. The curve is described by $V_{HT}^{(2)}=\frac{M_{2}^{(0)}}{l\Lambda_{c}}(lT)^{2}$. Right: $V_{H}^{(2)}$ as a function of $l\Lambda_{c}$ for fixed $lT=10$ in the high temperature limit. The curve is described by $V_{HT}^{(2)}=\frac{1}{l\Lambda_{c}}\frac{M_{1}^{(0)}}{lT}$.}
\label{fig4}
\end{figure}
where $B_1$ and $B_2$ are constant coefficients given in appendix \ref{B} and ${\rm{ln}}(4\epsilon)$ by considering the second order of $r_c/r_*$, is a function of $r_h$ and $l$ obtained in appendix B. Using \eqref{h13} and \eqref{h17} we have
\begin{align}
V_{HT}^{(2)}(l\Lambda_{c},lT)\equiv\frac{V_{\gamma_{A}}(l,l\Lambda_{c},lT)}{L^{2}\Lambda_{c}T}=\frac{1}{l^{2}\Lambda_{c}T}\bigg[(M_{1}^{(0)}+lTM_{2}^{(0)})(lT)^{2}+(M_{1}^{(2)}+lTM_{2}^{(2)})(l\Lambda_{c})^{2}\bigg],
\label{h18}
\end{align}
where $V_{HT}^{(2)}$ is dimensionless volume up to the $(r_c/r_*)^{2}$ at high temperature, $M_{1}^{(2)}$ and $M_{2}^{(2)}$ are
constant coefficients given by
\begin{align}
&M_{1}^{(2)}=B_1+B_2\bigg(\frac{\sqrt{6\pi}\Gamma(\frac{2}{3})}{3\Gamma(\frac{7}{6})}+\sum_{m=1}^{\infty}\bigg(\frac{\sqrt{6}\Gamma(m+\frac{1}{2})\Gamma(\frac{2m}{3}+\frac{2}{3})}{3\Gamma(m+1)\Gamma(\frac{2m}{3}+\frac{7}{6})}-\frac{1}{m}\bigg)\bigg)\cr &~~~~-A_{2}\sqrt{6}\sum_{n=0}^{\infty}\bigg[\frac{3(2n+1)\Gamma(n+\frac{1}{2})}{2(3n+2)\sqrt{\pi}\Gamma(n+1)}{_{2}F_{1}}\bigg(\frac{1}{2},\frac{3n}{2}+1;\frac{3n}{2}+2;1\bigg)\cr &~~~~~~~~~~~~~~~~~~~~~-\frac{(2n+1)\Gamma(n+\frac{1}{2})}{2(n+1)\sqrt{\pi}\Gamma(n+1)}{_{2}F_{1}}\bigg(\frac{1}{2},\frac{3n}{2}+\frac{3}{2};\frac{3n}{2}+\frac{5}{2};1\bigg)\bigg]\cr & M_{2}^{(2)}=-\pi\sqrt{6}B_{2}.
\label{h19}
\end{align}
The terms including $M_{1}^{(0)}$ and $M_{2}^{(0)}$ come from the order $(r_c/r_*)^{0}$ which appeared in equation \eqref{h10} and terms including $M_{1}^{(2)}$ and $M_{2}^{(2)}$ are corrections due to the non-conformal and thermal effects. Again HSC can be obtained using equations \eqref{CV} and \eqref{h18}. 
From equation \eqref{h18}, it is seen that if we fix $l\Lambda_{c}$ and increase $lT$, the $V_{HT}^{(2)}$ will increase. This can be better observed from the left panel of figure \ref{fig4}, where we plot  $V_{HT}^{(2)}$ as a function of $lT$ for fixed $l\Lambda_{c}$. This behavior is the opposite of the behavior of low temperature case, left panel of figure \ref{fig2}. On the other side, if one fixes $lT$ and increases $l\Lambda_{c}$, then the $V_{HT}^{(2)}$ will decrease, the right panel of figure \ref{fig4} where we plot $V_{HT}^{(2)}$ as a function of $l\Lambda_{c}$ for fixed $lT$. This behavior is the same as the behavior of low temperature case, right panel of figure \ref{fig2}. 

We would like to compare HSC in the low and high temperature limit dropping the non-conformal effect. To do so, we need to take the limit $l\Lambda_{c}\rightarrow 0$ of equation \eqref{h18} and \eqref{ccot8}. The former limit is just the equation \eqref{h10} and the later one has the following expression
\begin{align}
\frac{V_{\gamma_{A}}}{L^{2}T^{2}}=\frac{1}{(lT)^{2}}\bigg(\bar{w}_{1}+\bar{\omega}_{1}(lT)^{4}\bigg),
\label{h12}
\end{align}
where the left-hand side is the conformal dimensionless volume in the low temperature limit. Unlike equation \eqref{h10}, we do not obtain a linear relation in terms of $lT$ and it is observed that $lT$ has inverse effect on the mentioned subregion volume.

Note that, in section \ref{htv} and \ref{htc}, we use expression $r_{*}=r_h(1+\epsilon)$, $\epsilon\ll 1$ to investigate conveniently the high temperature behavior of the meson potential energy and HSC. To get analytical results, we do the calculations up to order ${\rm{ln}}(4\epsilon)$ for HSC, while the meson potential energy becomes zero up to that order.
\end{itemize}
\section{Conclusion}\label{dd6}
In this paper, we study zero temperature and finite temperature potential energy and HSC of a probe meson using AdS/CFT correspondence in a non-conformal model. We develop a systematic expansion for those non-local observables and focus on the high energy limit, $r_c\ll r_{*}$ or $l\Lambda_{c}\ll 1$, leading to the analytical expressions in low and high temperature limits. The length of quark and anti-quark pair is identified as the subregion length and we hence study the meson subregion complexity in the underlying model. In zero and low temperature limits, non-conformal corrections decrease meson potential energy and increase the absolute value of HSC. In low temperature limit, thermal corrections decrease meson potential energy and do not have a specific effect on HSC. But at leading order, non-conformal and thermal effects have an equal effect on meson potential energy and HSC, implying a decreasing impact on them. At high temperature, thermal effects increase HSC, while at leading order non-conformal effects decrease it. Furthermore, in the high temperature limit, meson potential energy becomes zero.  
Near the transition point our calculation shows that less information is needed to specify the meson state at zero temperature. Hence, one can say that the meson state at zero temperature is more desirable. Several problems arise which we leave for further investigations:
\begin{itemize}
\item It would be interesting to study analytically, by using such systematic expansion, the relation between meson potential and subregion complexity using action prescription for non-conformal theories and to compare it to the volume prescription. It is interesting to check whether such linear relation exists for action prescription.
\item There are other quantum information quantities such as purification and fidelity and one can analytically study their relations with meson bounding in non-conformal theories in order to establish the powerful properties of these quantities and match them with holographic conjectures in such non-conformal theories.

\end{itemize}

\appendix
\section{Potential energy}\label{A}
\begin{itemize}
\item{Zero temperature}: In subsection \ref{hev}, we reach equation \eqref{vv12} for the dimensionless meson potential energy $\tilde{V}_{q\bar{q}}$ at zero temperature in the high energy limit $r_{c}\ll r_{*}$. In the following we write the full calculations. Using equation \eqref{vv4} and keeping up to $(r_c/r_*)^{4}$, we obtain
\begin{align}
l(r_*)&=\frac{2}{r_*}\sum_{n=0}^{\infty}\frac{\Gamma(n+\frac{1}{2})}{\sqrt{\pi}\Gamma(n+1)}\int_{0}^{1}u^{4n+2}\bigg[1+(2n+1)(1-u^{2})\left(\frac{r_c}{r_*}\right)^{2}\cr &+\frac{1}{2}(2n+1)^{2}(1-u^2)^{2}\left(\frac{r_c}{r_*}\right)^{4} \bigg]du.
\label{v7}
\end{align}
Integrating the above equation, we have an expression for $l$ as a function of $r_{*}$
\begin{equation}
l(r_*)=\frac{2}{r_*}\bigg[a_1+a_2\left(\frac{r_c}{r_*}\right)^{2}+a_3\left(\frac{r_c}{r_*}\right)^{4} \bigg],\,\,\,\,\,\,\,\,\,\,\,
a_1,a_2,a_3>0,
\label{vv}
\end{equation} 
where constant coefficients are given by
\begin{align}
&a_1\equiv\sum_{n=0}^{\infty}\frac{\Gamma(n+\frac{1}{2})}{\sqrt{\pi}\Gamma(n+1)(4n+3)}=\frac{\sqrt{\pi}\Gamma(\frac{7}{4})}{3\Gamma(\frac{5}{4})},\cr 
&a_2\equiv\sum_{n=0}^{\infty}\frac{2\Gamma(n+\frac{1}{2})(2n+1)}{\sqrt{\pi}\Gamma(n+1)(4n+3)(4n+5)} \cr&~~~=\frac{2}{15}\sqrt{\frac{2}{\pi}}\bigg[5\Gamma\bigg(\frac{3}{4}\bigg)\Gamma\bigg(\frac{7}{4}\bigg)-\Gamma\bigg(\frac{1}{4}\bigg)\Gamma\bigg(\frac{9}{4}\bigg)\bigg]+\frac{2}{63}{_{3}F_{2}}\bigg(\frac{3}{2},\frac{7}{4},\frac{9}{4};\frac{11}{4},\frac{13}{4};1\bigg), \cr 
&a_3\equiv\sum_{n=0}^{\infty}\frac{4\Gamma(n+\frac{1}{2})(2n+1)^{2}}{\sqrt{\pi}\Gamma(n+1)(4n+3)(4n+5)(4n+7)}\cr&~~~=\frac{4\sqrt{\pi}}{15}\bigg(\frac{\Gamma(\frac{11}{4})}{\Gamma(\frac{5}{4})}-\frac{\Gamma(\frac{9}{4})}{\Gamma(\frac{3}{4})}\bigg)+\frac{4}{693}\bigg[2~{_{4}F_{3}}\bigg(\frac{4}{3},\frac{7}{4},2,\frac{9}{4};1,\frac{13}{4},\frac{15}{4};1\bigg)\cr &~~~ -7~{_{3}F_{2}}\bigg(\frac{3}{2},\frac{9}{4},\frac{11}{4};\frac{13}{4},\frac{15}{4};1\bigg) \bigg].
\label{coe1}
\end{align}
In high energy limit, the corrections to the pure AdS are small and then one can solve equation \eqref{vv} perturbatively for $r_*$
\begin{equation}
r_*(l)=\frac{1}{l}\bigg(\bar{a}_1+\frac{\bar{a}_2}{2}(l\Lambda_{c})^{2}+\frac{\bar{a}_3}{4}(l\Lambda_{c})^{4} \bigg)\,\,\,\,\,\,\,\,\bar{a}_{1},\bar{a}_{2}>0,\,\,\,\, \bar{a}_{3}<0,
\label{v9}
\end{equation} 
where $\bar{a}_{1}$, $\bar{a}_{2}$ and $\bar{a}_{3}$ are constant coefficients given by 
\begin{align}
&\bar{a}_1=\frac{2\sqrt{\pi}\Gamma(\frac{7}{4})}{3\Gamma(\frac{5}{4})},\cr& 
\bar{a}_2=\frac{1}{560\sqrt{2}\pi^{3/2}\Gamma(\frac{7}{4})^{2}}\bigg\lbrace315\pi^{2}+\Gamma\bigg(\frac{1}{4}\bigg)^{2}\bigg[-42~\Gamma\bigg(\frac{1}{4}\bigg)\Gamma\bigg(\frac{9}{4}\bigg)+5\sqrt{2\pi}~{_{3}F_{2}}\bigg(\frac{3}{2},\frac{7}{4},\frac{9}{4};\frac{11}{4},\frac{13}{4};1 \bigg)\bigg]\bigg\rbrace ,\cr &
\bar{a}_3=\frac{1}{574013440~\pi^{9/2}\Gamma(\frac{7}{4})^{4}}\bigg\lbrace-82657935360\sqrt{2}~\Gamma\bigg(\frac{5}{4}\bigg)^{10}+1968046080\sqrt{\pi}~\Gamma\bigg(\frac{1}{4}\bigg)\Gamma\bigg(\frac{5}{4}\bigg)^{7} \cr &~~~ {_{3}F_{2}}\bigg(\frac{3}{2},\frac{7}{4},\frac{9}{4};\frac{11}{4},\frac{13}{4}; 1 \bigg)+715\sqrt{\pi}~\bigg[\Gamma\bigg(\frac{1}{4}\bigg)^{6}\bigg(3969\pi-32~{_{3}F_{2}}\bigg(\frac{3}{2},\frac{7}{4},\frac{9}{4};\frac{11}{4},\frac{13}{4}; 1 \bigg)^{2}\bigg]
\cr &~~~ +2016\pi^{5/2}~\Gamma\bigg(\frac{1}{4}\bigg)^{4}\bigg[1430~{_{3}F_{2}}\bigg(\frac{3}{2},\frac{7}{4},\frac{9}{4};\frac{11}{4},\frac{13}{4}; 1 \bigg)  +910~{_{3}F_{2}}\bigg(\frac{3}{2},\frac{9}{4},\frac{11}{4};\frac{13}{4},\frac{15}{4}; 1 \bigg) \cr &~~~ -63~{_{3}F_{2}}\bigg(\frac{5}{2},\frac{11}{4},\frac{13}{4};\frac{17}{4},\frac{19}{4}; 1 \bigg) \bigg] \bigg\rbrace.
\label{coe2}
\end{align}
We do the same calculation for $\tilde{V}_{q\bar{q}}$, equation \eqref{v4}, and expanding it to $(r_{c}/r_{*})^{4}$ we get
\begin{align}
V_{q\bar{q}}(r_*)&=\frac{r_*}{\pi \alpha'}\sum_{n=0}^{\infty}\frac{\Gamma(n+\frac{1}{2})}{\sqrt{\pi}\Gamma(n+1)}\int_{\delta}^{1}u^{4n-2}\bigg[1+(2n+(1-2n)u^{2})\left(\frac{r_c}{r_*}\right)^{2} \cr & +\frac{1}{2}(2n+(1-2n)u^{2})^{2}\left(\frac{r_c}{r_*}\right)^{4}\bigg]du.
\label{v10}
\end{align} 
By integration of equation \eqref{v10} and considering finite terms, the terms which do not include $\delta$, we reach
\begin{equation}
V_{q\bar{q}}(r_*)=\frac{r_*}{\pi \alpha'}\bigg[b_1+b_2\left(\frac{r_c}{r_*}\right)^{2}+b_3\left(\frac{r_c}{r_*}\right)^{4} \bigg], \,\,\,\,\,\,\, b_1<0 ,\,\,\,\,\,b_2,b_3>0,
\label{v11}
\end{equation}
where constant coefficients $b_1$, $b_2$ and $b_3$ are given by
\begin{align}
&b_1\equiv\sum_{n=0}^{\infty}\frac{\Gamma(n+\frac{1}{2})}{\sqrt{\pi}\Gamma(n+1)(4n-1)}=-\frac{\sqrt{\pi}\Gamma(\frac{3}{4})}{\Gamma(\frac{1}{4})},\cr 
&b_2\equiv\sum_{n=0}^{\infty}\frac{\Gamma(n+\frac{1}{2})(1-8n)}{\sqrt{\pi}\Gamma(n+1)(1-16n^{2})}=\frac{\Gamma(\frac{1}{4})^{2}-\Gamma(-\frac{1}{4})\Gamma(\frac{3}{4})}{8\sqrt{2\pi}}+\frac{4}{15}{_{3}F_{2}}\bigg(\frac{3}{4},\frac{5}{4},\frac{3}{2};\frac{7}{4},\frac{9}{4};1\bigg),\cr &
b_3\equiv\sum_{n=0}^{\infty}\frac{\Gamma(n+\frac{1}{2})(1+8n-80n^{2})}{\sqrt{\pi}\Gamma(n+1)(6+8n-96n^{2}-128n^{3})} \cr&~~~ =\frac{\Gamma(\frac{5}{4})^{2}}{2\sqrt{2\pi}}+\frac{3\sqrt{\pi}\Gamma(\frac{7}{4})}{\Gamma(\frac{1}{4})}-\frac{9}{35}{_{3}F_{2}}\bigg(\frac{5}{4},\frac{3}{2},\frac{7}{4};\frac{9}{4},\frac{11}{4};1\bigg) \cr &~~~ +\frac{10}{231}{_{3}F_{2}}\bigg(\frac{7}{4},\frac{9}{4},\frac{5}{2};\frac{13}{4},\frac{15}{4};1\bigg) .
\label{coe3}
\end{align} 
Using equation \eqref{v9} and substituting $r_{c}=\Lambda_{c}/\sqrt{2}$, we reach the final result
\begin{equation}
\tilde{V}_{q\bar{q}}(l\Lambda_{c})\equiv\frac{\alpha' V_{q\bar{q}}(l,l\Lambda_{c})}{\Lambda_{c}}=\frac{1}{l\Lambda_{c}}\bigg(\bar{b}_{1}+\frac{\bar{b}_{2}}{2}(l\Lambda_{c})^{2}+\frac{\bar{b}_{3}}{4}(l\Lambda_{c})^{4} \bigg),\,\,\,\,\,\bar{b}_{1}<0,\,\,\,\, \bar{b}_{2}, \bar{b}_{3}>0,
\label{v13}
\end{equation} 
where $\tilde{V}_{q\bar{q}}$ is dimensionless meson potential energy at zero temperature. All the constant coefficients are summarized in table \ref{list1}. 
\begin{align}
\bar{b}_1&=-\frac{2\Gamma(\frac{3}{4})\Gamma(\frac{7}{4})}{3\Gamma(\frac{1}{4})\Gamma(\frac{5}{4})},\cr 
\bar{b}_2&=\frac{1}{8960\pi\Gamma(\frac{7}{4})^{2}}\bigg\lbrace   525~\Gamma\bigg(\frac{1}{4}\bigg)^{2}+315~\Gamma\bigg(\frac{-1}{4}\bigg)\Gamma\bigg(\frac{3}{4}\bigg)\cr & +16\sqrt{2\pi}\bigg[42~{_{3}F_{2}}\bigg(\frac{3}{4},\frac{5}{4},\frac{3}{2};\frac{7}{4},\frac{9}{4}; 1 \bigg) -5~{_{3}F_{2}}\bigg(\frac{3}{2},\frac{7}{4},\frac{9}{4};\frac{11}{4},\frac{13}{4} ; 1 \bigg)  \bigg]    \bigg\rbrace, \cr 
\bar{b}_3&=\frac{1}{4592107520~\pi^{4}~\Gamma(\frac{7}{4})^{4}}\bigg\lbrace\Gamma\bigg(\frac{1}{4}\bigg)^{2}\bigg[-630630\pi^{2}\bigg(87~\Gamma\bigg(\frac{1}{4}\bigg)^{2}-896~\Gamma\bigg(\frac{7}{4}\bigg)^{2}\bigg)\cr & +3003~\Gamma\bigg(\frac{1}{4}\bigg)^{4}\bigg[735~\Gamma\bigg(\frac{1}{4}\bigg)^{2}+8\sqrt{2\pi}\bigg(84~{_{3}F_{2}}\bigg(\frac{3}{4},\frac{5}{4},\frac{3}{2};\frac{7}{4},\frac{9}{4} ; 1 \bigg)-55~{_{3}F_{2}}\bigg(\frac{3}{2},\frac{7}{4},\frac{9}{4};\frac{11}{4},\frac{13}{4} ; 1 \bigg)\bigg)\bigg] \cr &-73216\pi~\Gamma\bigg(\frac{1}{4}\bigg)^{2}{_{3}F_{2}}\bigg(\frac{3}{2},\frac{7}{4},\frac{9}{4};\frac{11}{4},\frac{13}{4} ; 1 \bigg)\bigg[21~{_{3}F_{2}}\bigg(\frac{3}{4},\frac{5}{4},\frac{3}{2};\frac{7}{4},\frac{9}{4} ;1 \bigg)-5~{_{3}F_{2}}\bigg(\frac{3}{2},\frac{7}{4},\frac{9}{4};\frac{11}{4},\frac{13}{4} ; 1 \bigg)\bigg]\cr & -4032\sqrt{2}\pi^{5/2}\bigg[12012~{_{3}F_{2}}\bigg(\frac{3}{4},\frac{5}{4},\frac{3}{2};\frac{7}{4},\frac{9}{4} ; 1 \bigg)+11583~{_{3}F_{2}}\bigg(\frac{5}{4},\frac{3}{2},\frac{7}{4};\frac{9}{4},\frac{11}{4} ; 1 \bigg)\cr &-5005~{_{3}F_{2}}\bigg(\frac{3}{2},\frac{7}{4},\frac{9}{4};\frac{11}{4},\frac{13}{4} ;1 \bigg)+3640~{_{3}F_{2}}\bigg(\frac{3}{2},\frac{9}{4},\frac{5}{2};\frac{13}{4},\frac{15}{4}; 1 \bigg)+1950~{_{3}F_{2}}\bigg(\frac{7}{4},\frac{9}{4},\frac{5}{2};\frac{13}{4},\frac{15}{4} ; 1 \bigg)\cr &+252~{_{3}F_{2}}\bigg(\frac{5}{2},\frac{11}{4},\frac{13}{4};\frac{17}{4},\frac{19}{4}; 1 \bigg)  \bigg]          \bigg]             \bigg\rbrace.
\label{coe4}
\end{align}
\begin{table}[ht] 
\caption{Numerical constant coefficients for potential calculation}
\vspace{1 mm}
\centering
\begin{tabular}{ll}
\hline\hline
$a_1$= 0.59907 &\qquad\quad $a_2$= 0.355979 \quad\qquad $a_3$= 0.214869 \\

$\bar{a}_1$= 1.19814 &\qquad\quad $\bar{a}_2$= 0.495952 \quad\qquad $\bar{a}_3$= -0.202051 \\

$b_1$= -0.59907 & \qquad\quad $b_2$= 1.66701 \quad\qquad\, $b_3$= 0.813939 \\ 

$\bar{b}_1$= -0.228473 &\qquad\quad $\bar{b}_2$= 0.348301 \quad\qquad $\bar{b}_3$= 0.00584084 \\

\hline
\end{tabular}\\[1ex]
\label{list1}
\end{table}
\item{Low temperature:}
 In subsection \ref{ltv}, equation \eqref{vtt8} is reached for the dimensionless meson potential energy $\hat{V}_{q\bar{q}}$ at low temperature $lT\ll 1$ in the high energy limit $r_{c}\ll r_{*}$. In the following we review the details of computations.

Using \eqref{vt2} and expanding it up to $(r_c/r_*)^{4}$ and $(r_h/r_*)^{4}$ we get  
 \begin{align}
l(r_{*})&=\frac{2}{r_*}\sum_{n=0}^{\infty}\frac{\Gamma(n+\frac{1}{2})}{\sqrt{\pi}\Gamma(n+1)}\int_{0}^{1}u^{4n+2}\bigg\lbrace 1+\bigg((n+1)u^{4n+4}-(n+\frac{1}{2})\bigg)\left(\frac{r_h}{r_*}\right)^{4} \cr & +\bigg[(2n+1)(1-u^2)
+\bigg((2n+1)(1-u^2)((n+1)u^{4n+4}-(n+\frac{1}{2}))\bigg)\bigg(\frac{r_h}{r_*}\bigg)^{4}\bigg]\bigg(\frac{r_c}{r_*}\bigg)^{2}\cr & +\frac{1}{2}\bigg[(2n+1)^{2}(1-u^{2})^{2}+\bigg((2n+1)^{2}(1-u^{2})^{2}((n+1)u^{4n+4}-(n+\frac{1}{2}))\bigg)\bigg(\frac{r_h}{r_*}\bigg)^{4} \bigg]\cr &\bigg(\frac{r_c}{r_*}\bigg)^{4} \bigg\rbrace du.
\label{vt3}
\end{align}
Integrating equation \eqref{vt3}, we obtain $l$ in terms of $r_*$ 
 \begin{align}
l(r_{*})&=\frac{2}{r_*}\bigg[a_1+\alpha_{1}\bigg(\frac{r_h}{r_*}\bigg)^{4}+\bigg(a_2+\alpha_2 \bigg(\frac{r_h}{r_*}\bigg)^{4}\bigg)\bigg(\frac{r_c}{r_*}\bigg)^{2} \cr &+\bigg(a_3+\alpha_{3}\bigg(\frac{r_h}{r_*}\bigg)^{4}\bigg)\bigg(\frac{r_c}{r_*}\bigg)^{4} \bigg], \,\,\,\,\,\,\,\,\alpha_{1},\alpha_{2},\alpha_{3}<0,
\label{vt4}
\end{align}
where $\alpha_{1},\alpha_{2},\alpha_{3}$ are given by
\begin{align}
&\alpha_{1}\equiv-\sum_{n=0}^{\infty}\frac{\Gamma(n+\frac{1}{2})(4n+1)}{\sqrt{\pi}\Gamma(n+1)(32n^2+80n+42)}=-\frac{\sqrt{\pi}(11\Gamma(\frac{11}{4})+12\Gamma(\frac{15}{4}))}{924\Gamma(\frac{9}{4})},\cr 
&\alpha_{2}\equiv-\sum_{n=0}^{\infty}\frac{3\Gamma(n+\frac{1}{2})(2n+1)(16n^{2}+32n+11)}{\sqrt{\pi}\Gamma(n+1)(4n+3)(4n+5)(4n+7)(4n+9)}=\frac{\frac{1291}{84}\Gamma(\frac{5}{4})^{2}-\frac{71}{9}\Gamma(\frac{7}{4})^{2}}{\sqrt{2\pi}} \cr  &~~~-\frac{67}{286}~{_{3}F_{2}}\bigg(\frac{3}{2},\frac{11}{4},\frac{13}{4};\frac{15}{4},\frac{17}{4};1\bigg)-\frac{6}{187}~{_{3}F_{2}}\bigg(\frac{5}{2},\frac{11}{4},\frac{17}{4};\frac{19}{4},\frac{21}{4};1\bigg) \cr &~~~+\frac{6}{221}~{_{3}F_{2}}\bigg(\frac{5}{2},\frac{13}{4},\frac{15}{4};\frac{19}{4},\frac{21}{4};1\bigg)-\frac{16}{3003}~{_{5}F_{4}}\bigg(\frac{3}{2},\frac{7}{4},2,2,\frac{9}{4};1,1,\frac{15}{4},\frac{17}{4};1\bigg),
\cr &\alpha_{3}\equiv-\sum_{n=0}^{\infty}\frac{2\Gamma(n+\frac{1}{2})((2n+1)^{2}(80n^{2}+184n+69))}{\sqrt{\pi}\Gamma(n+1)(4n+3)(4n+5)(4n+7)(4n+9)(4n+11)} \cr&~~~ =-\frac{2}{135135}\bigg[897~{_{3}F_{2}}\bigg(\frac{1}{2},\frac{3}{4},\frac{5}{4};\frac{13}{4},\frac{15}{4};1\bigg)+230~{_{3}F_{2}}\bigg(\frac{3}{2},\frac{7}{4},\frac{9}{4};\frac{17}{4},\frac{19}{4};1\bigg) \cr &~~~+546~{_{4}F_{3}}\bigg(\frac{3}{2},\frac{7}{4},2,\frac{9}{4};1,\frac{17}{4},\frac{19}{4};1\bigg)+528~{_{5}F_{4}}\bigg(\frac{3}{2},\frac{7}{4},2,2,\frac{9}{4};1,1,\frac{17}{4},\frac{19}{4};1\bigg) \cr &~~~+160~{_{6}F_{5}}\bigg(\frac{3}{2},\frac{7}{4},2,2,2,\frac{9}{4};1,1,1,\frac{17}{4},\frac{19}{4};1\bigg)\bigg] .
\label{coet1}
\end{align}
In low temperature and high energy limits, the corrections to pure AdS and non- conformal terms are small and hence can be computed perturbatively. Solving equa- tion \eqref{vt4} order by order for $r_*$
\begin{align}
r_*(l)=\frac{1}{l}\bigg(\bar{a}_1+\frac{\bar{a}_2}{2}(l\Lambda_{c})^{2}+\frac{\bar{a}_3}{4}(l\Lambda_{c})^{4}+\bar{a}_{4}(lT)^{4} \bigg),\,\,\,\,\,\,\,\,\bar{a}_{1},\bar{a}_{2}>0,\,\,\,\, \bar{a}_{3},\bar{a}_{4}<0,
\label{vt5}
\end{align}
where $\bar{a}_{4}$ is given by
\begin{align}
\bar{a}_4=&-\frac{27\Gamma(\frac{5}{4})^{4}}{2464(\pi^{3/2}\Gamma(\frac{7}{4})^{4}\Gamma(\frac{9}{4}))}\bigg[11~\Gamma\bigg(\frac{11}{4}\bigg)+12~\Gamma\bigg(\frac{15}{4}\bigg)\bigg].
\end{align}
We use equation \eqref{vt33} and expand it up to $(r_c/r_*)^{4}$ and $(r_h/r_*)^{4}$. Then we have
\begin{align}
V_{q\bar{q}}(r_{*})&=\frac{r_*}{\pi\alpha'}\sum_{n=0}^{\infty}\frac{\Gamma(n+\frac{1}{2})}{\sqrt{\pi}\Gamma(n+1)}\int_{\delta}^{1}u^{4n-2}\bigg\lbrace 1+n(u^{4}-1)\bigg(\frac{r_h}{r_*}\bigg)^{4} \cr & +\bigg[2n+(1-2n)u^{2}+\bigg(n(2n+(1-2n)u^{2})(u^{4}-1)\bigg)\bigg(\frac{r_h}{r_*}\bigg)^{4}\bigg]\bigg(\frac{r_c}{r_*}\bigg)^{2}\cr & +\frac{1}{2}\bigg[(2n+(1-2n)u^{2})^{2}+\bigg(n(2n+(1-2n)u^{2})^{2}(u^{4}-1)\bigg)\bigg(\frac{r_h}{r_*}\bigg)^{4}\bigg] \cr & \bigg(\frac{r_c}{r_*}\bigg)^{4} \bigg\rbrace du.
\label{vt6}
\end{align}
Integrating equation \eqref{vt6} and keep the finite terms, terms without $\delta$ 
\begin{align}
V_{q\bar{q}}(r_{*})&=\frac{r_*}{\pi\alpha'}\bigg[b_1+\beta_{1}\bigg(\frac{r_h}{r_*}\bigg)^{4}+\bigg(b_2+\beta_{2}\bigg(\frac{r_h}{r_*}\bigg)^{4}\bigg)\bigg(\frac{r_c}{r_*}\bigg)^{2}\cr &+\bigg(b_{2}+\beta_{3}\bigg(\frac{r_h}{r_*}\bigg)^{4}\bigg)\bigg(\frac{r_c}{r_*}\bigg)^{4}\bigg],\,\,\,\,\,\,\,\beta_{1},\beta_{2},\beta_{3}<0
\label{vt7}
\end{align}
where $\beta_{1},\beta_{2}$ and $\beta_{3}$ are constant coefficients given by
\begin{align}
&\beta_{1}\equiv-\sum_{n=0}^{\infty}\frac{4n\Gamma(n+\frac{1}{2})}{\sqrt{\pi}\Gamma(n+1)(16n^{2}+8n-3)}=-\frac{2\sqrt{\pi}\Gamma(\frac{11}{4})}{21\Gamma(\frac{5}{4})},\cr 
&\beta_{2}\equiv-\sum_{n=0}^{\infty}\frac{12n\Gamma(n+\frac{1}{2})(16n^{2}+8n-1)}{\sqrt{\pi}\Gamma(n+1)(4n-1)(4n+1)(4n+3)(4n+5)}=\frac{5\Gamma(-\frac{1}{4})\Gamma(\frac{11}{4})-7\Gamma(-\frac{3}{4})\Gamma(\frac{13}{4})}{30\sqrt{2\pi}} \cr  &~~~-\frac{1}{9}~{_{3}F_{2}}\bigg(\frac{3}{2},\frac{7}{4},\frac{9}{4};\frac{11}{4},\frac{13}{4};1\bigg)-\frac{36{_{3}F_{2}}(\frac{7}{4},\frac{5}{2},\frac{13}{4};\frac{15}{4},\frac{17}{4};1)-28{_{3}F_{2}}(\frac{9}{4},\frac{5}{2},\frac{11}{4};\frac{15}{4},\frac{17}{4};1)}{1001} \cr &~~~-\frac{32}{315}~{_{5}F_{4}}\bigg(\frac{3}{4},\frac{5}{4},\frac{3}{2},2,2;1,1,\frac{11}{4},\frac{13}{4};1\bigg),
\cr& \beta_{3}\equiv-\sum_{n=0}^{\infty}\frac{2n\Gamma(n+\frac{1}{2})(4n(176n^{2}+140n-17)-5)}{\sqrt{\pi}\Gamma(n+1)(4n-1)(4n+1)(4n+3)(4n+5)(4n+7)} \cr&~~~ =-\frac{1}{10395}\bigg[5~{_{3}F_{2}}\bigg(\frac{3}{4},\frac{5}{4},\frac{3}{2};\frac{13}{4},\frac{15}{4};1\bigg)+68~{_{4}F_{3}}\bigg(\frac{3}{4},\frac{5}{4},\frac{3}{2},2;1,\frac{13}{4},\frac{15}{4};1\bigg) \cr &~~~-560~{_{5}F_{4}}\bigg(\frac{3}{4},\frac{5}{4},\frac{3}{2},2,2;1,1,\frac{13}{4},\frac{15}{4};1\bigg)\cr &~~~-704~{_{6}F_{5}}\bigg(\frac{3}{4},\frac{5}{4},\frac{3}{2},2,2,2;1,1,1,\frac{13}{4},\frac{15}{4};1\bigg)\bigg].
\label{coet2}
\end{align} 
Using equation \eqref{vt5} and substituting $r_{c}=\Lambda_{c}/\sqrt{2}$ and $r_h=\pi T$ we reach the final result
\begin{align}
\hat{V}_{q\bar{q}}(l\Lambda_{c},lT)\equiv\frac{\alpha'V_{q\bar{q}}(l,l\Lambda_{c},lT)}{\Lambda_{c}^{\frac{1}{2}}T^{\frac{1}{2}}}&=\frac{1}{l\Lambda_{c}^{\frac{1}{2}}T^{\frac{1}{2}}}\bigg[\bar{b}_{1}+\bar{\beta}_{1}(lT)^{4}+\frac{1}{2}(\bar{b}_{2}+\bar{\beta}_{2}(lT)^{4})(l\Lambda_{c})^{2}\cr&+\frac{1}{4}(\bar{b}_{3} +\bar{\beta}_{3}(lT)^{4})(l\Lambda_{c})^{4}\bigg],\,\,\,\,\,\bar{\beta}_{1}<0,\,\,\,\, \bar{\beta}_{2}, \bar{\beta}_{3}>0,
\label{vt8}
\end{align}
where $\hat{V}_{q\bar{q}}$ is dimensionless meson potential energy at low temperature. $\beta_{1},\beta_{2}$ and $\beta_{3}$ are constant coefficients which are given by
\begin{align}
\bar{\beta}_1&=-\frac{27\pi^{2}\Gamma(\frac{1}{4})\Gamma(\frac{5}{4})}{320\Gamma(\frac{7}{4})^{2}},\cr 
\bar{\beta}_2&=\frac{\pi^{2}}{3828825~\Gamma(-\frac{1}{4})^{7}}\bigg\lbrace 16~\Gamma\bigg(\frac{1}{4}\bigg)\bigg[312432120\sqrt{2}\pi^{2}-75810735\sqrt{2}\Gamma\bigg(\frac{1}{4}\bigg)^{4} \cr &-256\sqrt{\pi}\Gamma\bigg(\frac{1}{4}\bigg)^{2}\bigg(51051~{_{3}F_{2}}\bigg(\frac{3}{4},\frac{5}{4},\frac{3}{2};\frac{7}{4},\frac{9}{4};1\bigg)-303875~{_{3}F_{2}}\bigg(\frac{3}{2},\frac{7}{4},\frac{9}{4};\frac{11}{4},\frac{13}{4};1\bigg)\cr &+240975~{_{3}F_{2}}\bigg(\frac{7}{4},\frac{5}{2},\frac{13}{4};\frac{15}{4},\frac{17}{4};1\bigg)-187425~{_{3}F_{2}}\bigg(\frac{9}{4},\frac{5}{2},\frac{11}{4};\frac{15}{4},\frac{17}{4};1\bigg)\cr &+9450~{_{3}F_{2}}\bigg(\frac{11}{4},\frac{13}{4},\frac{7}{2};\frac{19}{4},\frac{21}{4};1\bigg)      \bigg) \bigg]\bigg\rbrace.
\label{coet3}
\end{align}
$\bar{\beta}_{3}$ is not reported here due to its complex form. The numerical value of all the constant coefficients are summarized in table \ref{listt1}.
\begin{table}[ht] 
\caption{Numerical constant coefficients for thermal potential calculation}
\vspace{1 mm}
\centering
\begin{tabular}{llll}
\hline\hline
$\alpha_{1}$= -0.119814 &\qquad\quad $\alpha_{2}$= -0.274949 \quad\qquad $\alpha_{3}$= -0.330873 \quad\qquad $\bar{a}_{4}$= -0.116281 \\

$\beta_{1}$= -0.299535 &\qquad\quad $\beta_{2}$= -0.671444 \quad\qquad $\beta_{3}$= -0.917564 \\

$\bar{\beta}_{1}$= -3.23986 &\qquad\quad $\bar{\beta}_{2}$= 2.4604 \quad\qquad\,\,\,\,\,\, $\bar{\beta}_{3}$= 1.94742 \\ 
\hline
\end{tabular}\\[1ex]
\label{listt1}
\end{table} 
\end{itemize}
\section{Holographic subregion complexity}\label{B}
\begin{itemize}
\item{Zero temperature:} In subsection \ref{hec}, we obtain equation \eqref{cco12} for the dimensionless subregion volume $\tilde{V}_{\gamma_{A}}$ at zero temperature in the high energy limit $r_c\ll r_*$. Note that $\tilde{V}_{\gamma_{A}}$ is applied to achieve HSC using equation \eqref{CV}. Here we review the computations.

Using equation \eqref{coo6} and up to $(r_c/r_*)^{4}$ and $(r_h/r_*)^{4}$ we get
\begin{align}
l(r_*)&=\frac{2}{r_*}\sum_{n=0}^{\infty}\frac{\Gamma(n+\frac{1}{2})}{\sqrt{\pi}\Gamma(n+1)}\int_{0}^{1}u^{6n+3}\bigg[1+\frac{3}{2}(2n+1)(1-u^{2})\bigg(\frac{r_c}{r_*}\bigg)^{2}\cr & +\frac{9}{8}(2n+1)^{2}(1-u^{2})^{2}\bigg(\frac{r_c}{r_*}\bigg)^{4}\bigg]du.
\label{co13}
\end{align}
Performing the above integral, we have
\begin{align}
l(r_*)=\frac{2}{r_*}\bigg[k_1+k_2\left(\frac{r_c}{r_*}\right)^{2}+k_3\left(\frac{r_c}{r_*}\right)^{4}  \bigg],\,\,\,\,k_1,k_2,k_3>0,
\label{co8}
\end{align}
where $k_{1}$, $k_2$ and $k_3$ are given by 
\begin{align}
&k_1\equiv\sum_{n=0}^{\infty}\frac{\Gamma(n+\frac{1}{2})}{\sqrt{\pi}\Gamma(n+1)(6n+4)}=\frac{\sqrt{\pi}\Gamma(\frac{5}{3})}{4\Gamma(\frac{7}{6})},\cr 
&k_2\equiv\sum_{n=0}^{\infty}\frac{\Gamma(n+\frac{1}{2})(1+2n)}{4\sqrt{\pi}\Gamma(n+1)(2+5n+3n^{2})}=\frac{1}{2}-\frac{\sqrt{\pi}\Gamma(\frac{5}{3})}{8\Gamma(\frac{7}{6})},\cr& 
k_3\equiv\sum_{n=0}^{\infty}\frac{3\Gamma(n+\frac{1}{2})(1+2n)^2}{8\sqrt{\pi}\Gamma(n+1)(8+26n+27n^{2}+9n^{3})} \cr &~~~ =\frac{3}{1120\pi}\bigg\lbrace 189\sqrt{\pi}\bigg[2\Gamma\bigg(\frac{5}{3}\bigg)\Gamma\bigg(\frac{11}{6}\bigg)+\Gamma\bigg(\frac{7}{6}\bigg)\Gamma\bigg(\frac{7}{3}\bigg)\bigg]\cr &~~~ +8\pi~{_{3}F_{2}}\bigg(\frac{3}{2},\frac{5}{3},\frac{7}{3};\frac{8}{3},\frac{10}{3};1\bigg)-280\pi  \bigg\rbrace.
\label{coe5}
\end{align}
Solving \eqref{co8} for $r_*$ perturbatively we get
\begin{align}
r_{*}(l)=\frac{1}{l}\bigg(\bar{k}_{1}+\frac{\bar{k}_{2}}{2}(\sqrt{c}l)^{2}+\frac{\bar{k}_{3}}{4}(\sqrt{c}l)^{4}\bigg),\,\,\,\,\bar{k}_1,\bar{k}_2>0,~~~~\bar{k}_3<0,
\label{co14}
\end{align}
where $\bar{k}_{1}$, $\bar{k}_{2}$ and $\bar{k}_{3}$ are given by
\begin{align}
&\bar{k}_1=\frac{\sqrt{\pi}\Gamma(\frac{5}{3})}{2\Gamma(\frac{7}{6})},\cr & 
\bar{k}_2=\frac{5\times2^{2/3}\pi\Gamma(\frac{4}{3})+4\Gamma(\frac{7}{6})^{2}(-5+2~{_{2}F_{1}}(1,\frac{4}{3},\frac{8}{3},-1))}{5\pi\Gamma(\frac{5}{3})^{2}},\cr &
\bar{k}_3=\frac{1}{25515\pi^{3/2}\Gamma(\frac{5}{3})^{5}\Gamma(\frac{11}{6})}\bigg\lbrace5600\times 2^{\frac{2}{3}}\sqrt{3}\pi^{\frac{5}{2}}\Gamma\bigg(\frac{1}{6}\bigg)-4725\times2^{1/3}\pi^{2}\Gamma\bigg(\frac{1}{3}\bigg)^{2}\cr &~~~+5670\times2^{2/3}\sqrt{\pi}\Gamma\bigg(\frac{1}{3}\bigg)^{2}\Gamma\bigg(\frac{7}{6}\bigg)^{3}+14~\Gamma\bigg(\frac{1}{6}\bigg)^{4}  \bigg[75-4\times 2^{2/3}~{_{2}F_{1}}\bigg(\frac{4}{3},\frac{5}{3};\frac{8}{3};-1\bigg)^{2}\bigg]\cr &~~~+5\times 2^{2/3}\pi~\Gamma\bigg(\frac{1}{6}\bigg)^{2}\Gamma\bigg(\frac{1}{3}\bigg)\bigg[-1085+3~{_{3}F_{2}}\bigg(\frac{3}{2},\frac{5}{3},\frac{7}{3};\frac{8}{3},\frac{10}{3}; 1 \bigg)\bigg]\bigg\rbrace.
\end{align}
Similar calculation for $V_{\gamma_{A}}$, equation \eqref{co6}, and keeping up to $(r_c/r_*)^{4}$ and $(r_h/r_*)^{4}$ we obtain
\begin{align}
\tilde{V}_{\gamma_{A}}(r_*)&=2L^{2}r_{*}^{2}\sum_{n=0}^{\infty}\frac{\Gamma(n+\frac{1}{2})}{\sqrt{\pi}\Gamma(n+1)}\int_{\delta}^{1}du\bigg\lbrace \frac{1-u^{6n+4}}{(6n+4)u^{4}}\cr &+\bigg[\frac{3}{2}(2n+1)\bigg(\frac{u^{-4}}{6(3n+2)(n+1)}+u^{6n}\bigg(\frac{u^{2}}{6(2n+1)}-\frac{1}{2(3n+2)}\bigg)\bigg)\cr&+\frac{u^{-2}}{3n+2}(1-u^{6n+4})\bigg]\bigg(\frac{r_c}{r_*}\bigg)^{2}+\bigg[3(2n+1)\bigg(\frac{u^{-2}}{6(3n+2)(n+1)}\cr &+u^{6n+2}\bigg(\frac{u^{2}}{6(n+1)}-\frac{1}{2(3n+2)}\bigg)\bigg)+\frac{1}{3n+2}(1-u^{6n+4})\cr &+\frac{9}{8}(2n+1)^{2}\bigg(\frac{u^{-4}}{3(3n+2)(3n+4)(n+1)}+u^{6n}\bigg(\frac{u^{2}}{3(n+1)}-\frac{u^{4}}{2(3n+4)}\cr & -\frac{1}{2(3n+2)}\bigg)\bigg)\bigg]\bigg(\frac{r_c}{r_*}\bigg)^{4} \bigg\rbrace.
\label{co15}
\end{align}
Integrating \eqref{co15} and considering finite parts
\begin{align}
V_{\gamma_{A}}(r_*)=2L^{2}r_{*}^{2}\bigg[w_{1}+w_{2}\bigg(\frac{r_c}{r_*}\bigg)^{2}+w_{3}\bigg(\frac{r_c}{r_*}\bigg)^{4}\bigg],\,\,\,\,\,w_{1}, w_{2}, w_{3}<0,
\label{co16}
\end{align}
where $w_1$, $w_2$ and $w_{3}$ are given by
\begin{align}
w_1\equiv&-\sum_{n=0}^{\infty}\frac{\Gamma(n+\frac{1}{2})}{3\sqrt{\pi}\Gamma(n+1)(6n+1)}=-\frac{\sqrt{\pi}\Gamma(\frac{7}{6})}{3\Gamma(\frac{2}{3})},\cr 
w_2\equiv&-\sum_{n=0}^{\infty}\frac{\Gamma(n+\frac{1}{2})(3+14n)}{\sqrt{\pi}\Gamma(n+1)(3+24n+36n^{2})}\cr =&\frac{\pi}{4}-\frac{\sqrt{\pi}\Gamma(\frac{7}{6})}{\Gamma(\frac{5}{3})}-\frac{1}{9}~{_{3}F_{2}}\bigg(\frac{7}{6},\frac{3}{2},\frac{3}{2};\frac{13}{6},\frac{5}{2};1\bigg),\cr 
w_3\equiv&-\sum_{n=0}^{\infty}\frac{\Gamma(n+\frac{1}{2})(1+2n)}{\sqrt{\pi}\Gamma(n+1)(5+36n+36n^{2})} \cr =&\frac{\sqrt{\pi}}{20}\bigg(\frac{\Gamma(\frac{11}{6})}{\Gamma(\frac{4}{3})}-\frac{5\Gamma(\frac{7}{6})}{\Gamma(\frac{2}{3})}\bigg)-\frac{1}{77}~{_{3}F_{2}}\bigg(\frac{7}{6},\frac{3}{2},\frac{11}{6};\frac{13}{6},\frac{17}{6};1\bigg).
\label{coe7}
\end{align}
Using equation \eqref{co15} and substituting $r_{c}=\Lambda_{c}/\sqrt{2}$ we reach the final result
\begin{equation}
\tilde{V}_{\gamma_{A}}(l\Lambda_{c})\equiv\frac{V_{\gamma_{A}}(l,l\Lambda_{c})}{L^2\Lambda_{c}^{2}}=\frac{1}{(l\Lambda_{c})^{2}}\bigg(\bar{w}_{1}+\frac{\bar{w}_{2}}{2}(l\Lambda_{c})^{2}+\frac{\bar{w}_{3}}{4}(l\Lambda_{c})^{4}  \bigg),\,\,\,\,\,\bar{w}_{1}, \bar{w}_{2}, \bar{w}_{3}<0,
\label{co12}
\end{equation}
where $\tilde{V}_{\gamma_{A}}$ is dimensionless subregion volume at zero temperature. $\bar{w}_{1}$,  $\bar{w}_{2}$ and $\bar{w}_{3}$ are given by
\begin{align}
&\bar{w}_1=-\frac{\pi^{3/2}\Gamma(\frac{5}{3})^{2}}{6\Gamma(\frac{2}{3})\Gamma(\frac{7}{6})},\cr & 
\bar{w}_2=\frac{1}{18}\bigg[9\pi-\frac{2\Gamma(\frac{1}{6})^{2}}{\Gamma(\frac{2}{3})^{2}}-\frac{7\sqrt{\pi}\Gamma(\frac{1}{6})}{\Gamma(\frac{2}{3})}-4~{_{3}F_{2}}\bigg(\frac{7}{6},\frac{3}{2},\frac{3}{2};\frac{13}{6},\frac{5}{2};1 \bigg) \bigg],\cr 
&\bar{w}_3=\frac{1}{1330560\times 2^{1/3}~\Gamma(\frac{2}{3})^{5}\Gamma(\frac{8}{3})}\bigg\lbrace\bigg[-1330560~2^{2/3}\sqrt{\pi}\Gamma\bigg(\frac{1}{3}\bigg)^{3}\Gamma\bigg(\frac{7}{6}\bigg)^{3}\cr &~~~-4065600\sqrt{3}~2^{2/3}\pi^{5/2}\Gamma\bigg(\frac{1}{6}\bigg)\Gamma\bigg(\frac{4}{3}\bigg)+693\pi^{2}\bigg[3200~2^{1/3}\Gamma\bigg(\frac{4}{3}\bigg)^{3}+243\Gamma\bigg(\frac{5}{3}\bigg)\Gamma\bigg(\frac{8}{3}\bigg)\Gamma\bigg(\frac{11}{3}\bigg)\bigg]\cr &~~~ +3193344~\Gamma\bigg(\frac{7}{6}\bigg)^{4}\Gamma\bigg(\frac{4}{3}\bigg)\bigg[-325+16\times 2^{2/3}~{_{2}F_{1}}\bigg(\frac{4}{3},\frac{5}{3};\frac{8}{3};-1\bigg)^{2}\bigg]\cr &~~~ -25600~2^{2/3}\pi^{3}~{_{3}F_{2}}\bigg(\frac{7}{6},\frac{3}{2},\frac{11}{6};\frac{13}{6},\frac{17}{6}; 1 \bigg) -42240~\pi~\Gamma\bigg(\frac{1}{3}\bigg)^{2}\Gamma\bigg(\frac{7}{6}\bigg)^{2}\bigg(84~{_{2}F_{1}}\bigg(\frac{4}{3},\frac{5}{3};\frac{8}{3};-1\bigg)\cr &~~~ +3\times2^{2/3}~{_{3}F_{2}}\bigg(\frac{3}{2},\frac{5}{3},\frac{7}{3};\frac{8}{3},\frac{10}{3}; 1 \bigg) -1330\times2^{2/3} \bigg) \bigg]\bigg\rbrace.
\label{coe8}
\end{align}
All the constant coefficients are summarized in the table \ref{list2}.
\begin{table}[ht] 
\caption{Numerical constant coefficients for potential calculation}
\vspace{1 mm}
\centering
\begin{tabular}{ll}
\hline\hline
$k_1$= 0.431185 &\qquad\quad $k_2$= 0.284408 \quad\qquad $k_3$= 0.180262 \\

$\bar{k}_1$=0.86237 &\qquad\quad $\bar{k}_2$= 0.764864 \quad\qquad $\bar{k}_3$= -0.704896 \\

$w_1$= -0.404775 &\qquad\quad $w_2$= -1.45197 ~~~~~~~~ $w_3$= -0.264624 \\ 

$\bar{w}_1$= -0.602048 &\qquad\quad $\bar{w}_2$= -3.9719 ~~~~~~~~~  $\bar{w}_3$= -0.201038 \\

\hline
\end{tabular}\\[1ex]
\label{list2}
\end{table}
\item{Low temperature:} In subsection \ref{ltc}, equation \eqref{ccot8} is reached for the dimensionless subregion volume $\hat{V}_{\gamma_{A}}$ at low temperature $lT\ll 1$ in the high energy limit $r_c\ll r_*$. HSC is obtained using equation \eqref{CV}. In the following we write the full calculation here.

Expanding equation \eqref{cot22} up to $(r_c/r_*)^{4}$ and $(r_h/r_*)^4$ we have
\begin{align}
l(r_*)&=\frac{2}{r_*}\sum_{n=0}^{\infty}\frac{\Gamma(n+\frac{1}{2})}{\sqrt{\pi}\Gamma(n+1)}\int_{0}^{1}u^{6n+3}\bigg\lbrace 1+\frac{1}{2}u^{6n+4}\bigg(\frac{r_h}{r_*}\bigg)^{4} \cr & +\bigg[\frac{3}{2}(2n+1)(1-u^{2})+\frac{3}{4}(2n+1)(1-u^2)u^{6n+4}\bigg(\frac{r_h}{r_*}\bigg)^{4}\bigg]\bigg(\frac{r_c}{r_*}\bigg)^{2} \cr &
+\bigg[\frac{9}{8}(2n+1)^{2}(1-u^{2})^{2}+\frac{9}{18}(2n+1)^{2}(1-u^2)^{2}u^{6n+4}\bigg(\frac{r_h}{r_*}\bigg)^{4}\bigg]\cr &\bigg(\frac{r_c}{r_*}\bigg)^{4} \bigg\rbrace du.
\label{cot3}
\end{align}
Integrating equation \eqref{cot3} we obtain
\begin{align}
l(r_*)&=\frac{2}{r_*}\bigg[k_1+\kappa_{1}\bigg(\frac{r_h}{r_*}\bigg)^{4}+\bigg(k_2+\kappa_{2}\bigg(\frac{r_h}{r_*}\bigg)^{4}\bigg)\bigg(\frac{r_c}{r_*}\bigg)^{2}+\cr &\bigg(k_3+\kappa_{3}\bigg(\frac{r_h}{r_*}\bigg)^{4}\bigg)\bigg(\frac{r_c}{r_*}\bigg)^{4} \bigg],\,\,\,\,\,\,\kappa_{1},\kappa_{2},\kappa_{3}>0
\label{cot4}
\end{align}
where $\kappa_{1},\kappa_{2},\kappa_{3}$ are given by
\begin{align}
\kappa_{1}\equiv&\sum_{n=0}^{\infty}\frac{\Gamma(n+\frac{1}{2})}{\sqrt{\pi}\Gamma(n+1)(12n+16)}=\frac{\sqrt{\pi}\Gamma(\frac{7}{3})}{16\Gamma(\frac{11}{6})},\cr 
\kappa_{2}\equiv&\sum_{n=0}^{\infty}\frac{\Gamma(n+\frac{1}{2})(6n+3)}{\sqrt{\pi}\Gamma(n+1)(72n^{2}+216n+160)}\cr =&\frac{3}{2240}\bigg[\frac{72(3\Gamma(\frac{13}{6})\Gamma(\frac{7}{3})-2\Gamma(\frac{5}{6})\Gamma(\frac{8}{3}))}{\sqrt{\pi}}+5~{_{3}F_{2}}\bigg(\frac{3}{2},\frac{7}{3},\frac{8}{3};\frac{10}{3},\frac{11}{3};1\bigg)\bigg],
\cr \kappa_{3}\equiv&\sum_{n=0}^{\infty}\frac{\Gamma(n+\frac{1}{2})(2n+1)^{2}}{6\sqrt{\pi}\Gamma(n+1)(9n^{3}+45n^{2}+74n+40)} \cr =&1+\frac{9(95\Gamma(\frac{13}{6})\Gamma(\frac{7}{3})-184\Gamma(\frac{11}{6})\Gamma(\frac{8}{3}))}{700\sqrt{\pi}}+\frac{1}{168}~{_{3}F_{2}}\bigg(\frac{3}{2},\frac{7}{3},\frac{8}{3};\frac{10}{3},\frac{11}{3};1\bigg).
\label{coet4}
\end{align} 
Solving equation \eqref{cot4} perturbatively we reach
\begin{equation}
r_*(l)=\frac{1}{l}\bigg(\bar{k}_{1}+\frac{\bar{k}_2}{2}(\sqrt{c} l)^{2}+\frac{\bar{k}_3}{4}(\sqrt{c} l)^{4} +\bar{k}_{4}(r_h l)^{4} \bigg), \,\,\,\,\,\bar{k}_1, \bar{k}_2,\bar{k}_{4}>0,\,\,\,\,\,\,\,\bar{k}_3<0,
\label{cot5}
\end{equation}
where $\bar{k}_{4}$ is given by
\begin{align}
\bar{k}_4=&\frac{2\Gamma(\frac{7}{6})^{4}\Gamma(\frac{7}{3})}{\pi^{3/2}\Gamma(\frac{5}{3})^{4}\Gamma(\frac{11}{6})}.
\label{coe6}
\end{align}
Similar calculations yield to an expression for $V_{\gamma_{A}}$ from equation \eqref{cot33}
\begin{align}
V_{\gamma_{A}}(r_*)&=2V_{y,z}r_*^2\sum_{n=0}^{\infty}\frac{\Gamma(n+\frac{1}{2})}{\sqrt{\pi}\Gamma(n+1)} \int_{\delta}^{1}du\bigg\lbrace\frac{1-u^{6n+4}}{(6n+4)u^{4}}+\bigg(\frac{1-u^{6n+8}}{2(6n+8)u^{4}}+\frac{1-u^{6n+4}}{2(6n+4)}\bigg)\bigg(\frac{r_h}{r_*}\bigg)^{4} \cr & +\bigg[\frac{3}{2}(2n+1)\bigg(\frac{u^{-4}}{6(3n+2)(n+1)}+u^{6n}\bigg(\frac{u^2}{6(n+1)}-\frac{1}{2(3n+2)}\bigg)\bigg)+\frac{u^{-2}}{3n+2}(1-u^{6n+4}) \cr & 
+\bigg(\frac{3}{4}(2n+1)\bigg(\frac{u^{-4}}{2(3n+4)(3n+5)}+u^{6n}\bigg(\frac{(3n+4)u^{6}}{3(3n+5)(n+1)}+\frac{u^{4}}{(3n+2)(3n+4)}\bigg)\cr &+\frac{1}{6(3n+2)(n+1)}\bigg)+\frac{u^{-2}}{2(3n+4)}(1-u^{6n+8})+\frac{u^{2}}{2(3n+2)}(1-u^{6n+4})\bigg)\bigg(\frac{r_h}{r_*}\bigg)^{4}
\bigg]\bigg(\frac{r_c}{r_*}\bigg)^{2}\cr & +\bigg[3(2n+1)\bigg(\frac{u^{-2}}{6(3n+2)(n+1)}+u^{6n+2}\bigg(\frac{u^{2}}{6(n+1)}-\frac{1}{2(3n+2)}\bigg) \bigg)+\frac{1}{3n+2}(1-u^{6n+4}) \cr & +\frac{9}{8}(2n+1)^{2}\bigg(\frac{u^{-4}}{3(3n+2)(3n+4)(n+1)}+u^{6n}\bigg(\frac{u^{2}}{3(n+1)}-\frac{u^4}{2(3n+4)}-\frac{1}{2(3n+2)}\bigg)\bigg) \cr & +\bigg(\frac{3}{2}(2n+1)\bigg(\frac{u^{-2}}{2(3n+4)(3n+5)}+\frac{u^{2}}{6(3n+2)(n+1)}+\frac{3u^{6n+6}(n+1)}{(3n+2)(3n+4)}+\cr &\frac{u^{6n+8}(3n+4)}{3(3n+5)(n+1)}\bigg)+\frac{9}{16}(2n+1)^2\bigg(\frac{u^{-4}}{3(3n+5)(3n+4)(n+2)}+\frac{1}{(3n+4)(3n+2)(n+1)}\cr & +u^{6n}\bigg(\frac{2(3n+4)u^{6}}{3(3n+5)(n+1)u^{6}}-\frac{3(n+1)u^{4}}{(3n+2)(3n+4)}-\frac{(6n+5)u^{8}}{6(3n+4)(n+2)}\bigg)\bigg)+\frac{u^{4}(1-u^{6n+4})}{6n+4} \cr & +\frac{1}{6n+8}(1-u^{6n+8})\bigg)\bigg(\frac{r_h}{r_*}\bigg)^{4} \bigg]\bigg(\frac{r_c}{r_*}\bigg)^{4} \bigg\rbrace du.
\label{cot6}
\end{align}
Integrating \eqref{cot6} and considering finite parts yield
\begin{align}
V_{\gamma_{A}}(r_*)&=2V_{y,z}r_*^2\bigg[w_1+\omega_{1}\bigg(\frac{r_h}{r_*}\bigg)^{4}+\bigg(w_{2}+\omega_{2}\bigg(\frac{r_h}{r_*}\bigg)^{4}\bigg)\bigg(\frac{r_c}{r_*}\bigg)^{2}\cr &+ \bigg(w_3+\omega_{3}\bigg(\frac{r_h}{r_*}\bigg)^{4}\bigg)\bigg(\frac{r_c}{r_*}\bigg)^{4} \bigg],\,\,\,\,\, \omega_{1},\omega_{3}>0\,\,\,\,, \omega_{2}<0,
\label{cot7}
\end{align} 
where $\omega_{1},\omega_{3}$ and $\omega_{2}$ are given by
\begin{align}
\omega_{1}\equiv&\sum_{n=0}^{\infty}\frac{\Gamma(n+\frac{1}{2})}{\sqrt{\pi}\Gamma(n+1)(18n+15)}=\frac{\sqrt{\pi}\Gamma(\frac{11}{6})}{15\Gamma(\frac{4}{3})},\cr 
\omega_{2}\equiv&-\sum_{n=0}^{\infty}\frac{\Gamma(n+\frac{1}{2})}{\sqrt{\pi}\Gamma(n+1)(18n+15)}=-\frac{\sqrt{\pi}\Gamma(\frac{11}{6})}{15\Gamma(\frac{4}{3})},
\cr \omega_{3}\equiv&\sum_{n=0}^{\infty}\frac{\Gamma(n+\frac{1}{2})(26n+25)}{15\sqrt{\pi}\Gamma(n+1)(2n+3)(6n+5)} \cr =&\frac{5\pi}{24}-\frac{1}{6}~{_{3}F_{2}}\bigg(\frac{5}{6},\frac{3}{2},\frac{3}{2};\frac{11}{6},\frac{5}{2};1\bigg)+\frac{13}{825}~{_{3}F_{2}}\bigg(\frac{3}{2},\frac{11}{6},\frac{5}{2};\frac{17}{6},\frac{7}{2};1\bigg).
\label{coet5}
\end{align}
Then by substituting \eqref{cot5} in \eqref{cot7} we get
\begin{align} 
\hat{V}_{\gamma_{A}}(l\Lambda_{c},lT)\equiv\frac{V_{\gamma_{A}}(l,l\Lambda_{c},lT)}{L^{2}\Lambda_{c}T}&=\frac{1}{l^{2}\Lambda_{c}T}\bigg(\bar{w}_{1}+\bar{\omega}_{1}(lT)^{4}+\frac{1}{2}(\bar{w}_2+\bar{\omega}_{2}(lT)^{4})(l\Lambda_{c})^{2}\cr &+\frac{1}{4}(\bar{w}_3+\bar{\omega}_{3}(lT)^{4})(l\Lambda_{c})^{4}\bigg),\,\,\,\,\bar{\omega}_{1},\bar{\omega}_{2}<0\,\,\,,\bar{\omega}_{3}>0,
\label{cot8}
\end{align}
where $\hat{V}_{\gamma_{A}}$ is dimensionless subregion volume and $\bar{\omega}_{1}$, $\bar{\omega}_{2}$ and $\bar{\omega}_{3}$ are given by
\begin{align}
&\bar{\omega}_1=\frac{\pi^{7/2}\Gamma(\frac{7}{6})}{\sqrt{3}\Gamma(\frac{5}{3})}-\frac{108\times2^{1/3}\pi^{2}\Gamma(\frac{7}{6})^{6}}{5\Gamma(\frac{2}{3})^{3}},\cr 
&\bar{\omega}_2=-\frac{2^{4/3}\pi^{3/2}\Gamma(\frac{7}{6})^{2}}{9\Gamma(\frac{2}{3})^{7}\Gamma(\frac{8}{3})}\bigg[10\sqrt{3}\pi^{7/2}-9\pi\Gamma\bigg(\frac{1}{3}\bigg)^{3}\Gamma\bigg(\frac{7}{6}\bigg)^{3}+162\times2^{1/3}\Gamma\bigg(\frac{1}{3}\bigg)^{2}\Gamma\bigg(\frac{7}{6}\bigg)^{5}\bigg],\cr & \bar{\omega}_3=\frac{\pi}{35925120\times2^{1/3}\Gamma(\frac{2}{3})^{10}\Gamma(\frac{11}{6})\Gamma(\frac{8}{3})}\bigg\lbrace27720000\times2^{1/3}\pi^{6}\Gamma\bigg(\frac{1}{6}\bigg)-115500\pi^{3}\Gamma\bigg(\frac{1}{6}\bigg)^{7} \cr &~~~-96096000\sqrt{3}\pi^{11/2}\Gamma\bigg(\frac{1}{3}\bigg)^{2}+59675\times2^{2/3}\pi\Gamma\bigg(\frac{1}{6}\bigg)^{7}\Gamma\bigg(\frac{1}{3}\bigg)^{3} \cr &~~~+404157600\times2^{1/3}\pi^{2}\Gamma\bigg(\frac{1}{3}\bigg)^{4}\Gamma\bigg(\frac{7}{6}\bigg)^{5}-484989120\times2^{2/3}\sqrt{\pi}\Gamma\bigg(\frac{1}{3}\bigg)^{4}\Gamma\bigg(\frac{7}{6}\bigg)^{8} \cr &~~~ +154\Gamma\bigg(\frac{1}{6}\bigg)^{9}\Gamma\bigg(\frac{1}{3}\bigg)^{2}\bigg[2^{8/3}{_{2}F_{1}}\bigg(\frac{4}{3},\frac{5}{3};\frac{8}{3};-1\bigg)^{2}-75\bigg]\cr &~~~ +46200\pi^{7/2}\Gamma\bigg(\frac{1}{6}\bigg)^{4}\bigg[\sqrt{3}\bigg(8~{_{2}F_{1}}\bigg(\frac{4}{3},\frac{5}{3};\frac{8}{3};-1\bigg)+15\times2^{2/3}\bigg)-2^{8/3}\pi\bigg] \cr &~~~ +7200\pi^{2}\Gamma\bigg(\frac{1}{6}\bigg)^{3}\Gamma\bigg(\frac{1}{3}\bigg)^{5}{_{3}F_{2}}\bigg(\frac{7}{6},\frac{3}{2},\frac{11}{6};\frac{13}{6},\frac{17}{6};1\bigg) \cr &~~~-110\sqrt{3}\pi^{3/2}\Gamma\bigg(\frac{1}{6}\bigg)^{8}\Gamma\bigg(\frac{1}{3}\bigg){_{3}F_{2}}\bigg(\frac{3}{2},\frac{5}{3},\frac{7}{3};\frac{8}{3},\frac{10}{3};1\bigg) \cr &~~~ + 240\times2^{1/3}\pi^{9/2}\Gamma\bigg(\frac{1}{6}\bigg)^{2}\Gamma\bigg(\frac{1}{3}\bigg)\bigg[77000\sqrt{3}+9625\pi-12320\sqrt{3}~{_{2}F_{1}}\bigg(1,\frac{4}{3};\frac{8}{3};-1\bigg) \cr &~~~ -7700~{_{3}F_{2}}\bigg(\frac{5}{6},\frac{3}{2},\frac{3}{2};\frac{11}{6},\frac{5}{2};1\bigg)-330\sqrt{3}~{_{3}F_{2}}\bigg(\frac{3}{2},\frac{5}{3},\frac{7}{3};\frac{8}{3},\frac{10}{3};1\bigg) \cr &~~~ + 728~{_{3}F_{2}}\bigg(\frac{3}{2},\frac{11}{6},\frac{5}{2};\frac{17}{6},\frac{7}{2};1\bigg)  \bigg]   \bigg\rbrace.
\label{coet6}
\end{align}
All constant coefficients are reported in table \ref{listt2}.
\begin{table}[ht] 
\caption{Numerical constant of thermal subregion complexity }
\vspace{1 mm}
\centering
\begin{tabular}{llll}
\hline\hline
$\kappa_{1}$= 0.140218 &\qquad\quad $\kappa_{2}$= 0.0806394 \quad\qquad $\kappa_{3}$= 0.0480308 \quad\qquad $\bar{k}_{4}$= 0.507061 \\

$\omega_{1}$= 0.124472 &\qquad\quad $\omega_{2}$= -0.124472 \quad\qquad $\omega_{3}$= 0.245496 \\

$\bar{\omega}_{1}$= -36.3574 & \qquad\quad $\bar{\omega}_{2}$= -162.853 \quad\qquad\, $\bar{\omega}_{3}$= 539.984 \\ 
\hline
\end{tabular}\\[1ex]
\label{listt2}
\end{table}
\item{High temperature:} The constant coefficients corresponding to the equations \eqref{h9} and \eqref{h17} in subsection \ref{htc} are given as follows
\begin{align}
A_{1}&=\frac{\sqrt{\pi}\Gamma(\frac{1}{6})}{18\Gamma(\frac{2}{3})}{_{5}F_{4}}\bigg(-\frac{1}{4},\frac{1}{12},\frac{1}{6},\frac{1}{2},\frac{7}{12};\frac{1}{3},\frac{1}{3},\frac{2}{3},\frac{3}{4};1\bigg) \cr &-\frac{\sqrt{\pi}\Gamma(\frac{5}{6})}{12\Gamma(\frac{4}{3})}{_{5}F_{4}}\bigg(\frac{1}{12},\frac{5}{12},\frac{1}{2},\frac{5}{6},\frac{11}{12};\frac{2}{3},\frac{13}{12},\frac{4}{3},\frac{3}{4};1\bigg)\cr & -\frac{\pi}{160}{_{5}F_{4}}\bigg(\frac{5}{12},\frac{3}{4},\frac{5}{6},\frac{7}{6},\frac{5}{4};1,\frac{4}{3},\frac{17}{12},\frac{5}{3};1\bigg)\cr &-{_{2}F_{1}}\bigg(-\frac{3}{4},\frac{1}{2};\frac{1}{4};1\bigg)\bigg[\sum_{m=1}^{\infty}\bigg(\frac{\Gamma(m+\frac{1}{2})\Gamma(\frac{2m}{3}+\frac{2}{3})}{18\Gamma(m+1)\Gamma(\frac{2m}{3}+\frac{7}{6})}-\frac{1}{18m}\bigg)+\frac{\sqrt{\pi}\Gamma(\frac{2}{3})}{18\Gamma(\frac{7}{6})}\bigg]\cr & +\sum_{m=1}^{\infty}\bigg[\frac{\Gamma(m+\frac{1}{2})\Gamma(\frac{2m}{3}+\frac{1}{6})}{18\Gamma(m+1)\Gamma(\frac{2m}{3}+\frac{2}{3})}{_{6}F_{5}}\bigg(-\frac{1}{4},\frac{1}{6},\frac{1}{2},\frac{5}{6},\frac{m}{3}+\frac{1}{12},\frac{m}{3}+\frac{7}{12};\frac{1}{3},\frac{2}{3},\frac{3}{4},\frac{m}{3}+\frac{1}{3},\frac{m}{3}+\frac{5}{6};1\bigg)\cr & -\frac{\sqrt{3}}{18\sqrt{2}}{_{4}F_{3}}\bigg(-\frac{1}{4},\frac{1}{6},\frac{1}{2},\frac{5}{6};\frac{1}{3},\frac{2}{3},\frac{3}{4};1\bigg)m^{-1}\bigg]-\sum_{m=1}^{\infty}\bigg[\frac{\Gamma(m+\frac{1}{2})\Gamma(\frac{2m}{3}+\frac{5}{6})}{12\Gamma(m+1)\Gamma(\frac{2m}{3}+\frac{4}{3})}\cr &{_{6}F_{5}}\bigg(\frac{1}{12},\frac{1}{2},\frac{5}{6},\frac{7}{6},\frac{m}{3}+\frac{5}{12},\frac{m}{3}+\frac{11}{12};\frac{2}{3},\frac{13}{12},\frac{4}{3},\frac{m}{3}+\frac{2}{3},\frac{m}{3}+\frac{7}{6};1\bigg)\cr & -\frac{\sqrt{3}\Gamma(\frac{1}{6})\Gamma(\frac{13}{12})\Gamma(\frac{4}{3})}{2\sqrt{2}\Gamma(\frac{1}{12})\Gamma(\frac{1}{3})\Gamma(\frac{7}{6})}~{_{4}F_{3}}\bigg(\frac{1}{12},\frac{1}{2},\frac{5}{6},\frac{7}{6};\frac{2}{3},\frac{13}{12},\frac{4}{3};1\bigg)m^{-1}\bigg]-\sum_{m=1}^{\infty}\bigg[\frac{\Gamma(m+\frac{1}{2})\Gamma(\frac{2m}{3}+\frac{3}{2})}{80\Gamma(m+1)\Gamma(\frac{2m}{3}+2)}\cr & {_{6}F_{5}}\bigg(\frac{5}{12},\frac{5}{6},\frac{7}{6},\frac{3}{2},\frac{m}{3}+\frac{3}{4},\frac{m}{3}+\frac{5}{4};\frac{4}{3},\frac{17}{12},\frac{5}{3},\frac{m}{3}+1,\frac{m}{3}+\frac{3}{2};1\bigg)\cr & -\frac{9\sqrt{3}\Gamma(\frac{1}{6})\Gamma(\frac{4}{3})\Gamma(\frac{17}{12})\Gamma(\frac{5}{3})}{400\sqrt{2}\Gamma(\frac{1}{3})\Gamma(\frac{5}{12})\Gamma(\frac{2}{3})\Gamma(\frac{7}{6})}~{_{4}F_{3}}\bigg(\frac{5}{12},\frac{5}{6},\frac{7}{6},\frac{3}{2};\frac{4}{3},\frac{17}{12},\frac{5}{3};1\bigg)m^{-1} \bigg],
\label{c2}
\end{align}
\begin{align}
A_{2}&=\frac{\sqrt{3}\Gamma(\frac{1}{6})\Gamma(\frac{13}{12})\Gamma(\frac{4}{3})}{2\sqrt{2}\Gamma(\frac{1}{12})\Gamma(\frac{1}{3})\Gamma(\frac{7}{6})}~{_{4}F_{3}}\bigg(\frac{1}{12},\frac{1}{2},\frac{5}{6},\frac{7}{6};\frac{2}{3},\frac{13}{12},\frac{4}{3};1\bigg)+\frac{9\sqrt{3}\Gamma(\frac{1}{6})\Gamma(\frac{4}{3})\Gamma(\frac{17}{12})\Gamma(\frac{5}{3})}{400\sqrt{2}\Gamma(\frac{1}{3})\Gamma(\frac{5}{12})\Gamma(\frac{2}{3})\Gamma(\frac{7}{6})}\cr &{_{4}F_{3}}\bigg(\frac{5}{12},\frac{5}{6},\frac{7}{6},\frac{3}{2};\frac{4}{3},\frac{17}{12},\frac{5}{3};1\bigg)-\frac{\sqrt{3}}{18\sqrt{2}}{_{4}F_{3}}\bigg(-\frac{1}{4},\frac{1}{6},\frac{1}{2},\frac{5}{6};\frac{1}{3},\frac{2}{3},\frac{3}{4};1\bigg) \cr &+\frac{\sqrt{3}}{18\sqrt{2}}{_{2}F_{1}}\bigg(-\frac{3}{4},\frac{1}{2};\frac{1}{4};1\bigg).
\label{c3}
\end{align}
\begin{align}
B_{1}&=\sum_{n=0}^{\infty}\sum_{m=0}^{\infty}\bigg[\frac{3(2n+1)\Gamma(n+\frac{1}{2})\Gamma(m+\frac{1}{2})}{4\pi(3n+2m+3)(6n+4m+3)\Gamma(m+1)\Gamma(n+1)}{_{2}F_{1}}\bigg(\frac{1}{2},m+\frac{3n}{2}+\frac{3}{4};m+\frac{3n}{2}+\frac{7}{4};1\bigg)\cr & ~~~~~~~~~~~-\frac{3(2n+1)\Gamma(n+\frac{1}{2})\Gamma(m+\frac{1}{2})}{4\pi(3n+2m+2)(6n+4m+1)\Gamma(n+1)\Gamma(m+1)}{_{2}F_{1}}\bigg(\frac{1}{2},m+\frac{3n}{2}+\frac{1}{4};m+\frac{3n}{2}+\frac{5}{4};1\bigg)   \bigg]\cr & -\frac{\pi}{3}~{_{5}F_{4}}\bigg(-\frac{1}{12},\frac{1}{6},\frac{1}{4},\frac{3}{4},\frac{5}{6};\frac{1}{3},\frac{2}{3},\frac{11}{12},1;1\bigg)+\frac{\sqrt{\pi}\Gamma(\frac{13}{6})}{21\Gamma(\frac{5}{3})}~{_{5}F_{4}}\bigg(\frac{1}{4},\frac{1}{2},\frac{7}{12},\frac{13}{12},\frac{7}{6};\frac{2}{3},\frac{5}{4},\frac{4}{3},\frac{4}{3};1\bigg)\cr & +\frac{3\sqrt{\pi}\Gamma(\frac{17}{6})}{308\Gamma(\frac{7}{3})}~{_{5}F_{4}}\bigg(\frac{7}{12},\frac{5}{6},\frac{11}{12},\frac{17}{12},\frac{3}{2};\frac{4}{3},\frac{19}{12},\frac{5}{3},\frac{5}{3};1\bigg)\cr & -\sum_{m=1}^{\infty}\bigg(\frac{\Gamma(m+\frac{1}{2})\Gamma(\frac{2m}{3}+\frac{3}{2})}{(4m+3)(4m+7)(4m+11)\Gamma(m+1)\Gamma(\frac{2m}{3}+1)}(154+144m)\bigg)\cr&~~~~~~~~~~~{_{6}F_{5}}\bigg(-\frac{1}{12},\frac{1}{6},\frac{1}{2},\frac{5}{6},\frac{m}{3}+\frac{1}{4},\frac{m}{3}+\frac{3}{4};\frac{1}{3},\frac{2}{3},\frac{11}{12},\frac{m}{3}+\frac{1}{2},\frac{m}{3}+1;1\bigg)\cr &  +\sum_{m=1}^{\infty}\bigg(\frac{\Gamma(m+\frac{1}{2})\Gamma(\frac{2m}{3}+\frac{13}{6})}{(4m+3)(4m+7)(4m+11)\Gamma(m+1)\Gamma(\frac{2m}{3}+\frac{5}{3})}(11+\frac{56}{3}m)\bigg)\cr&~~~~~~~~~~{_{6}F_{5}}\bigg(\frac{1}{4},\frac{1}{2},\frac{5}{6},\frac{7}{6},\frac{m}{3}+\frac{7}{12},\frac{m}{3}+\frac{13}{12};\frac{2}{3},\frac{5}{4},\frac{4}{3},\frac{m}{3}+\frac{5}{6},\frac{m}{3}+\frac{4}{3};1\bigg)\cr & +\sum_{m=1}^{\infty}\bigg(\frac{\Gamma(m+\frac{1}{2})\Gamma(\frac{2m}{3}+\frac{17}{6})}{(4m+3)(4m+7)(4m+11)\Gamma(m+1)\Gamma(\frac{2m}{3}+\frac{7}{3})}(\frac{9}{4}+\frac{30}{7}m)\bigg)\cr&~~~~~~~~~~~{_{6}F_{5}}\bigg(\frac{7}{12},\frac{5}{6},\frac{7}{6},\frac{3}{2},\frac{m}{3}+\frac{11}{12},\frac{m}{3}+\frac{17}{12};\frac{4}{3},\frac{19}{12},\frac{5}{3},\frac{m}{3}+\frac{7}{6},\frac{m}{3}+\frac{5}{3};1\bigg)\cr & -\sum_{m=1}^{\infty}\bigg(\frac{32\Gamma(m+\frac{1}{2})\Gamma(\frac{2m}{3}+\frac{3}{2})m^{2}}{(4m+3)(4m+7)(4m+11)\Gamma(m+1)\Gamma(\frac{2m}{3}+1)}\cr&~~~~~~~~~~~~{_{6}F_{5}}\bigg(-\frac{1}{12},\frac{1}{6},\frac{1}{2},\frac{5}{6},\frac{m}{3}+\frac{1}{4},\frac{m}{3}+\frac{3}{4};\frac{1}{3},\frac{2}{3},\frac{11}{12},\frac{m}{3}+\frac{1}{2},\frac{m}{3}+1;1\bigg)\cr &~~~~~~~~~~~-\frac{1}{\sqrt{6}}{_{4}F_{3}}\bigg(-\frac{1}{12},\frac{1}{6},\frac{1}{2},\frac{5}{6};\frac{1}{3},\frac{2}{3},\frac{11}{12};1\bigg)m^{-1}\bigg)\cr & +\sum_{m=1}^{\infty}\bigg(\frac{16\Gamma(m+\frac{1}{2})\Gamma(\frac{2m}{3}+\frac{13}{6})m^{2}}{3(4m+3)(4m+7)(4m+11)\Gamma(m+1)\Gamma(\frac{2m}{3}+\frac{5}{3})}\cr&~~~~~~~~~~~~{_{6}F_{5}}\bigg(\frac{1}{4},\frac{1}{2},\frac{5}{6},\frac{7}{6},\frac{m}{3}+\frac{7}{12},\frac{m}{3}+\frac{13}{12};\frac{2}{3},\frac{5}{4},\frac{4}{3},\frac{m}{3}+\frac{5}{6},\frac{m}{3}+\frac{4}{3};1\bigg)\cr &~~~~~~~~~~~-\frac{\sqrt{2}\Gamma(\frac{1}{6})\Gamma(\frac{5}{4})\Gamma(\frac{4}{3})}{6\sqrt{3}\Gamma(\frac{1}{4})\Gamma(\frac{1}{3})\Gamma(\frac{7}{6})}~{_{4}F_{3}}\bigg(\frac{1}{4},\frac{1}{2},\frac{5}{6},\frac{7}{6};\frac{2}{3},\frac{5}{4},\frac{4}{3};1\bigg)m^{-1}\bigg)\cr & +\sum_{m=1}^{\infty}\bigg(\frac{12\Gamma(m+\frac{1}{2})\Gamma(\frac{2m}{3}+\frac{17}{6})m^{2}}{7(4m+3)(4m+7)(4m+11)\Gamma(m+1)\Gamma(\frac{2m}{3}+\frac{7}{3})}\cr&~~~~~~~~~~~~{_{6}F_{5}}\bigg(\frac{7}{12},\frac{5}{6},\frac{7}{6},\frac{3}{2},\frac{m}{3}+\frac{11}{12},\frac{m}{3}+\frac{17}{12};\frac{4}{3},\frac{19}{12},\frac{5}{3},\frac{m}{3}+\frac{7}{6},\frac{m}{3}+\frac{5}{3};1\bigg)\cr &~~~~~~~~~~~-\frac{27\sqrt{2}\Gamma(\frac{1}{6})\Gamma(\frac{4}{3})\Gamma(\frac{19}{12})\Gamma(\frac{5}{3})}{784\sqrt{3}\Gamma(\frac{1}{3})\Gamma(\frac{7}{12})\Gamma(\frac{2}{3})\Gamma(\frac{7}{6})}~{_{4}F_{3}}\bigg(\frac{7}{12},\frac{5}{6},\frac{7}{6},\frac{3}{2};\frac{4}{3},\frac{19}{12},\frac{5}{3};1\bigg)m^{-1}\bigg) ,
\label{c4}
\end{align}
\begin{align}
B_{2}&=\frac{1}{\sqrt{6}}~{_{4}F_{3}}\bigg(-\frac{1}{12},\frac{1}{6},\frac{1}{2},\frac{5}{6};\frac{1}{3},\frac{2}{3},\frac{11}{12};1\bigg)-\frac{\sqrt{2}\Gamma(\frac{1}{6})\Gamma(\frac{5}{4})\Gamma(\frac{4}{3})}{6\sqrt{3}\Gamma(\frac{1}{4})\Gamma(\frac{1}{3})\Gamma(\frac{7}{6})}~{_{4}F_{3}}\bigg(\frac{1}{4},\frac{1}{2},\frac{5}{6},\frac{7}{6};\frac{2}{3},\frac{5}{4},\frac{4}{3};1\bigg)\cr & -\frac{27\sqrt{2}\Gamma(\frac{1}{6})\Gamma(\frac{4}{3})\Gamma(\frac{19}{12})\Gamma(\frac{5}{3})}{784\sqrt{3}\Gamma(\frac{1}{3})\Gamma(\frac{7}{12})\Gamma(\frac{2}{3})\Gamma(\frac{7}{6})}~{_{4}F_{3}}\bigg(\frac{7}{12},\frac{5}{6},\frac{7}{6},\frac{3}{2};\frac{4}{3},\frac{19}{12},\frac{5}{3};1\bigg).
\label{c5}
\end{align}
Now for finding ${\rm{ln}}(4\epsilon)$ as a function of $l$ at leading and sub-leading order, we follow below procedure: At high temperature, we reach the finite expression for $l$ up to $(r_c/r_*)^2$, equation \eqref{h1}. Here, we consider zero order of $(r_c/r_*)$ and do sum over $n$, we obtain
\begin{align}
l(r_*)=\frac{1}{r_*}\sum_{m=0}^{\infty}\frac{\Gamma(m+\frac{1}{2})\Gamma(\frac{2m}{3}+\frac{2}{3})}{3\Gamma(m+1)\Gamma(\frac{2m}{3}+\frac{7}{6})}\bigg(\frac{r_h}{r_*}\bigg)^{4m}.
\label{d2}
\end{align}
For large $m$, this series goes as $\sim{m^{-1}(r_h/r_*)^{4m}}$ and hence it diverges for $r_*=r_h$. By isolating the divergence part of the series \eqref{d2}, we get
\begin{align}
l(r_*)&=\frac{1}{r_*}\bigg[\frac{\sqrt{\pi}\Gamma(\frac{2}{3})}{3\Gamma(\frac{7}{6})}+\sum_{m=1}^{\infty}\bigg(\frac{\Gamma(m+\frac{1}{2})\Gamma(\frac{2m}{3}+\frac{2}{3})}{3\Gamma(m+1)\Gamma(\frac{2m}{3}+\frac{7}{6})}-\frac{1}{\sqrt{6} m}\bigg)\bigg(\frac{r_h}{r_*}\bigg)^{4m}\cr &-\frac{1}{\sqrt{6}}{\rm{ln}}\bigg[1-\bigg(\frac{r_h}{r_*}\bigg)^{4}\bigg]\bigg].
\label{d3}
\end{align}
The convergence of the above infinite series can be obviously seen for $r_*=r_h$. We write $r_*=r_h(1+\epsilon)$, following the fact that at high temperature $\epsilon\ll 1$, we obtain
\begin{align}
{\rm{ln}}(4\epsilon)=-\sqrt{6}r_{h}l+\frac{\sqrt{6\pi}\Gamma(\frac{2}{3})}{3\Gamma(\frac{7}{6})}+\sum_{m=1}^{\infty}\bigg(\frac{\sqrt{6}\Gamma(m+\frac{1}{2})\Gamma(\frac{2m}{3}+\frac{2}{3})}{3\Gamma(m+1)\Gamma(\frac{2m}{3}+\frac{7}{6})}-\frac{1}{m}\bigg)+{\cal{O}}(\epsilon).
\label{d4}
\end{align}
Now, we consider the second order of $(r_c/r_*)$ and sum over $n$ for zero order of $(r_c/r_*)$
and over $m$ for second order of $(r_c/r_*)$, we get
\begin{align}
l(r_*)&=\frac{1}{r_*}\sum_{m=0}^{\infty}\frac{\Gamma(m+\frac{1}{2})\Gamma(\frac{2m}{3}+\frac{2}{3})}{3\Gamma(m+1)\Gamma(\frac{2m}{3}+\frac{7}{6})}\bigg(\frac{r_h}{r_*}\bigg)^{4m}\cr & +\frac{1}{r_*}\sum_{n=0}^{\infty}\bigg[\frac{3(2n+1)\Gamma(n+\frac{1}{2})}{2(3n+2)\sqrt{\pi}\Gamma(n+1)}{_{2}F_{1}}\bigg(\frac{1}{2},\frac{3n}{2}+1;\frac{3n}{2}+2;\bigg(\frac{r_h}{r_*}\bigg)^{4}\bigg)\cr &~~~~~~~~~~~ -\frac{(2n+1)\Gamma(n+\frac{1}{2})}{2(n+1)\sqrt{\pi}\Gamma(n+1)}{_{2}F_{1}}\bigg(\frac{1}{2},\frac{3n}{2}+\frac{3}{2};\frac{3n}{2}+\frac{5}{2};\bigg(\frac{r_h}{r_*}\bigg)^{4}\bigg)\cr &~~~~~~~~~~~\bigg]\bigg(\frac{r_c}{r_*}\bigg)^{2}.
\label{e2}
\end{align}
The investigation of convergence of the first sum, summation over $m$, is similar to the equation \eqref{d2} which was studied before. For large $n$, the behavior of two terms corresponding to the second sum, summation over $n$, is the same and hence it is a convergent series. We write $r_*=r_h(1+\epsilon)$, knowing that at high temperature $\epsilon\ll 1$, we get
\begin{align}
{\rm{ln}}(4\epsilon)&=-\sqrt{6}r_{h}l+\frac{\sqrt{6\pi}\Gamma(\frac{2}{3})}{3\Gamma(\frac{7}{6})}+\sum_{m=1}^{\infty}\bigg(\frac{\sqrt{6}\Gamma(m+\frac{1}{2})\Gamma(\frac{2m}{3}+\frac{2}{3})}{3\Gamma(m+1)\Gamma(\frac{2m}{3}+\frac{7}{6})}-\frac{1}{m}\bigg)\cr &+\sqrt{6}\sum_{n=0}^{\infty}\bigg[\frac{3(2n+1)\Gamma(n+\frac{1}{2})}{2(3n+2)\sqrt{\pi}\Gamma(n+1)}{_{2}F_{1}}\bigg(\frac{1}{2},\frac{3n}{2}+1;\frac{3n}{2}+2;1\bigg)\cr &~~~~~~~~~~~~ -\frac{(2n+1)\Gamma(n+\frac{1}{2})}{2(n+1)\sqrt{\pi}\Gamma(n+1)}{_{2}F_{1}}\bigg(\frac{1}{2},\frac{3n}{2}+\frac{3}{2};\frac{3n}{2}+\frac{5}{2};1\bigg)\bigg]\bigg(\frac{r_c}{r_h}\bigg)^{2}\cr &+{\cal{O}}(\epsilon). 
\label{e3}
\end{align}
As expected when $r_c=0$, we have \eqref{d4}.
\end{itemize}

\end{document}